\documentclass[twocolumn,english,aps,superscriptaddress,prl]{revtex4-1}

\usepackage{babel}
\usepackage{amsmath}
\usepackage{amssymb}
\usepackage{graphicx}
\usepackage{ytableau}

\usepackage{hyperref}
\usepackage{color}
\usepackage{mathrsfs}

\usepackage{comment}

\usepackage{xcolor}
\usepackage{tikz}
\usetikzlibrary{decorations}
\usetikzlibrary{decorations.pathmorphing}
\usetikzlibrary{decorations.pathreplacing}
\usetikzlibrary{decorations.markings}
\usetikzlibrary{shapes.misc}
\usetikzlibrary{calc}

\newcommand{\SU}[1]{{{\rm SU}({#1})}}
\newcommand{\irrepsymm}{[3\,0\,0]}
\newcommand{\irrepadj}{[2\,1\,0]}
\newcommand{\irrepsix}{[6\,0\,0]}
\newcommand{\irrepconj}{[3\,3\,0]}

\newcommand{\shsc}[3]{\mathrel{\raisebox{#1}{\protect\scalebox{#2}{#3}}}}

\hypersetup{
    colorlinks=true,
    linkcolor=blue,
    citecolor=blue,    
    urlcolor=blue,
}
\begin{document}

\title{Haldane Gap of the Three-Box Symmetric $\SU{3}$ Chain}

\author{Samuel Gozel}
\affiliation{Institute of Physics, Ecole Polytechnique F\'ed\'erale de Lausanne (EPFL), CH-1015 Lausanne, Switzerland}
\author{Pierre Nataf}
\affiliation{Laboratoire de Physique et Mod\'elisation des Milieux Condens\'es, Universit\'e Grenoble Alpes and CNRS, 25 avenue des Martyrs, 38042 Grenoble, France}
\author{Fr\'ed\'eric Mila}
\affiliation{Institute of Physics, Ecole Polytechnique F\'ed\'erale de Lausanne (EPFL), CH-1015 Lausanne, Switzerland}

\date{\today}
\begin{abstract} 
Motivated by the recent generalization of the Haldane conjecture to $\SU{3}$ chains [M. Lajk\'o {\it et al.}, Nucl. Phys. B\textbf{924}, 508 (2017)] according to which a Haldane gap should be present for symmetric representations if the number of boxes in the Young diagram is a multiple of three, we develop a density matrix renormalization group algorithm based on standard Young tableaus to study the model with three boxes directly in the representations of the global $\SU{3}$ symmetry. We show that there is a finite gap between the singlet and the symmetric $\irrepsymm$ sector  $\Delta_{\irrepsymm}/J = 0.040\pm0.006$ where $J$ is the antiferromagnetic Heisenberg coupling, and we argue on the basis of the structure of the low energy states that this is sufficient to conclude that the spectrum is gapped.
\end{abstract}

\maketitle


The role played in the field of quantum magnetism by Haldane's prediction that integer spin chains have a gap while half-odd-integer spin chains do not can hardly be overemphasized~\citep{haldane_continuum_1983,haldane_nonlinear_1983}. Indeed, ever since this prediction was confirmed experimentally and numerically, the community has thought differently about antiferromagnets, and the actual value of the spin (and not simply its length as a measure of quantum fluctuations) has become a defining parameter along with the dimensionality of space, the topology of the lattice and the isotropy of the couplings~\cite{renard_quantum_1988,
regnault_wave_1992,
white_density_1992,
white_density_1993,
white_numerical_1993,
golinelli_finite_1994,
todo_cluster_2001,
todo_parallel_2019,
nakano_haldane_2019}. 
With the progress made in ultracold fermionic experiments~\cite{
wu_exact_2003,
honerkamp_ultrcold_2004,
ultracold_cazalilla_2009,
gorshkov_two_2010,
taie_su6_2012,
nonne_symmetry_2013,
pagano_one_2014,
scazza_observation_2014,
zhang_spectroscopic_2014,
hofrichter_direct_2016,
capponi_phases_2016,
weichselbaum_unified_2018}, the $\SU{N}$ cousins of the $\SU{2}$ Heisenberg model have become the focus of a lot of attention, and quite naturally the question of whether and how Haldane's conjecture can be generalized has been addressed by many authors~\cite{
affleck_exact_1986,
affleck_critical_1988,
bykov_geometry_2013,
lajko_generalization_2017,
tanizaki_anomaly_2018,
wamer_selfconjugate_2019,
wamer_generalization_2020,
wamer_mass_2020}.
The generalization of the Affleck-Kennedy-Lieb-Tasaki (AKLT) construction has proven to be a very useful guide in predicting which model may or may not be gapped, as well as the concept of spinon confinement~\cite{affleck_rigorous_1987,
affleck_valence_1988,
greiter_exact_2007,
rachel_spinon_2009,
morimoto_z_2014,
wamer_selfconjugate_2019}.
More recently, Haldane's semiclassical derivation of a nonlinear sigma model with a topological term has been generalized to $\SU{3}$ chains in the symmetric representation, and the absence of topological terms in the action when the number of boxes in the Young diagram is a multiple of three is a strong indication that there should be a Haldane gap in that case without any symmetry breaking~\cite{lajko_generalization_2017}. The underlying theory, the $\SU{3}/[{\rm U}(1) \times {\rm U}(1)]$ flag manifold nonlinear sigma model, is of interest in itself as a nontrivial generalization of the standard $\mathbb{CP}^2$ model. For instance, it has been shown using 't Hooft anomaly matching that, when the number of boxes in the Young diagram is not a multiple of three, the model is gapless in the IR and is described by a $\SU{3}_1$ Wess-Zumino-Witten (WZW) conformal field theory~\cite{ohmori_sigma_2019,
tanizaki_anomaly_2018}.


On the numerical front, the presence of a finite gap in the three-box symmetric $\SU{3}$ chain, the simplest possible case where the Lieb-Schultz-Mattis-Affleck theorem does not apply (or equivalently no 't Hooft anomaly is present in the flag manifold nonlinear sigma model), is not at all clear.
In early density matrix renormalization group (DMRG) simulations, the saturation of the entanglement entropy has been interpreted as the evidence of a Haldane gap~\cite{rachel_spinon_2009}. This conclusion has been challenged, however, by exact diagonalizations (ED) which showed that, if there is a gap, the correlation length associated with it must be much larger than that at which the entanglement saturates according to Ref.~\cite{rachel_spinon_2009}, suggesting that the saturation of the entanglement entropy was actually a consequence of the truncation of the Hilbert space in DMRG~\cite{nataf_exact_2016}. To actually have a chance to detect the gap, one must clearly study much longer chains, and keep far more states. This is a real challenge because it is not known {\it a priori} how long the chain will have to be, and how many states will have to be kept. As for $\SU{2}$ with spin $S$, this length scale is expected to increase exponentially with the number of boxes in the Young diagram, but since the gap has never been calculated for any irreducible representation (irrep), the prefactor is not known, and an estimate of the length is not available.

In this Letter, we have taken on this challenge, and have developed a DMRG code in the basis of standard Young tableaus (SYTs) that allows one to take advantage of the full $\SU{N}$ symmetry, and to keep the equivalent of a huge number of states without increasing too much the size of the variational space~\cite{white_density_1992,white_density_1993}. 
This has allowed us to obtain definitive numerical evidence that the spectrum is gapped, and to come up with an estimate $\Delta_{\irrepsymm}/J\simeq 0.04$ for the gap in the $\irrepsymm$ sector, where $J$ is the antiferromagnetic coupling. Given the extremely small value of this gap, comparable to that of $\SU{2}$ chains for spin $S=3$ for which the correlation length $\xi \simeq  637$, it is clear {\it a posteriori} that there was no chance to detect it in the early attempts with standard DMRG or ED, and that it will be very challenging to look at cases with a larger number of boxes~\cite{todo_cluster_2001}.

The Hamiltonian of antiferromagnetic $\SU{N}$ Heisenberg chains can be written as
\begin{equation}
\label{eq:su3-hw-H}
\mathcal{H} = J \sum_{i} \sum_{\alpha,\beta=1}^{N} \mathcal{S}^{\alpha\beta}_i \mathcal{S}^{\beta\alpha}_{i+1}, \quad J>0
\end{equation}
where the generators $\mathcal{S}^{\alpha\beta}$ satisfy the usual $\SU{N}$ commutation relations
$[ \mathcal{S}^{\alpha\beta}, \mathcal{S}^{\mu\nu}] = \delta^{\mu\beta} \mathcal{S}^{\alpha\nu} - \delta^{\alpha\nu} \mathcal{S}^{\mu\beta}$.
For the ten-dimensional symmetric irrep of $\SU{3}$ represented by a Young diagram with three boxes in the first row, $\shsc{0pt}{0.4}{\ydiagram{3}} \ \equiv \irrepsymm$~\footnote{We denote an irrep $\alpha$ of $\SU{3}$ by $3$ integers $[\alpha_1 \, \alpha_2 \, \alpha_3]$ corresponding to the lengths of the rows in the corresponding Young diagram. It will indeed sometimes be useful to keep track of the columns with $3$ boxes.}, it can be written equivalently as
\begin{equation}
\label{eq:su3-hw-H}
\mathcal{H} = 2 J \sum_{i} \bold{T}_i \cdot \bold{T}_{i+1}
\end{equation}
where $T_i^a, \ a=1,...,8$ are ten-dimensional Hermitian traceless matrices representing the generators of $\mathfrak{su}(3)$ and are the exact analogues of the $\mathfrak{su}(2)$ spin operators $S^x$, $S^y$ and $S^z$ (see Supplemental Material~\cite{supp_mat} for details).

\begin{figure}
\begin{center}
\includegraphics[width=0.5\textwidth]{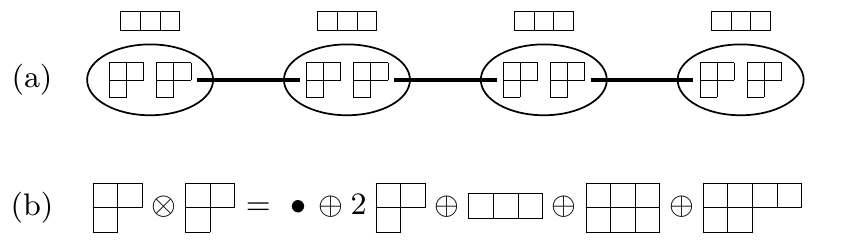}
\end{center}
\caption{(a) $\SU{3}$ AKLT state for the physical three-box symmetric irrep at each site. An ellipse denotes the projection of two adjoint representations onto the physical three-box symmetric irrep. The thick lines joining neighboring sites represent singlets made out of two adjoint irreps. (b) Decomposition of the tensor product of two adjoint representations.}
\label{fig::aklt_state_su3}
\end{figure}

To reach long enough chains, the only option is to use DMRG with open boundary conditions. In this case, edge states are expected to be present however. This is best understood by looking at the AKLT version of the model with a biquadratic interaction~\cite{supp_mat},
\begin{equation}
\mathcal{H}_{\rm AKLT} = 5 J \sum_i \left( \bold{T}_i \cdot \bold{T}_{i+1} + \frac{1}{5} (\bold{T}_i \cdot \bold{T}_{i+1})^2 + \frac{6}{5} \right),
\label{eq:su3_H_aklt}
\end{equation}
for which an exact ground state can be constructed~\cite{greiter_valence_2007}. There are different ways of constructing this wave function, but the most economical one consists in writing the irrep $\irrepsymm$ as a symmetrized product of two eight-dimensional adjoint irreps $\irrepadj$, and to make singlets with adjoint representations on neighboring sites, as illustrated in Fig.~\ref{fig::aklt_state_su3}(a)~\cite{gozel_novel_2019}. This implies that there are edge states in the adjoint representation. Accordingly, the spectrum of a finite chain will have low-lying states corresponding to all the representations appearing in the product of two adjoint representations and given in Fig.~\ref{fig::aklt_state_su3}(b). This is inconvenient for two reasons: the bulk gap does not correspond to the first excited state, and  the coupling between the edge states creates long-range entanglement that makes the convergence of DMRG much more difficult. To overcome this problem, and following what has been done for $\SU{2}$ spin chains, we have added an adjoint representation at each end of the chain, with a positive coupling to ensure that it forms a singlet with the edge state~\cite{white_numerical_1993,
schollwock_haldane_1995,
schollwock_s2antiferro_1996,
wang_haldane_1999,
tatsuaki_interaction_2000}. The precise value of the coupling is not important and the simulations have been done with a coupling equal to $J$~\cite{supp_mat}.

The DMRG code we have developed is an extension of the code used by two of the present authors to study $\SU{N}$ chains in the fundamental representation and which builds on previous developments for exact diagonalization~\cite{nataf_exact_2014,nataf_exact_2016,wan_exact_2017,nataf_density_2018}. It is based on the formulation of the Hamiltonian in terms of permutation operators~\cite{young_onsubstitutional_1900,supp_mat} and on the basis of SYTs. This code does not require knowing the $\SU{N}$ Clebsch-Gordan coefficients but relies on the subduction coefficients of the symmetric group~\cite{supp_mat,chen_grouptheory}. We are currently limited to the calculation of the subduction coefficients when the outer multiplicity is one \cite{nataf_density_2018}.
Technically this means that we cannot diagonalize the Hamiltonian in sectors characterized by certain $\SU{N}$ symmetry, such as the adjoint representation $\irrepadj$, to which the first excited state belongs according to ED on small chains~\cite{supp_mat}. We can, however, diagonalize the Hamiltonian in the singlet sector, which is the irrep of the ground state whatever the number of sites thanks to the adjoint edge irreps, and also in the symmetric irrep $\irrepsymm$~\footnote{One can also work in the $\irrepconj$ sector.}. 

\begin{figure}
\begin{center}
\includegraphics[width=0.5\textwidth]{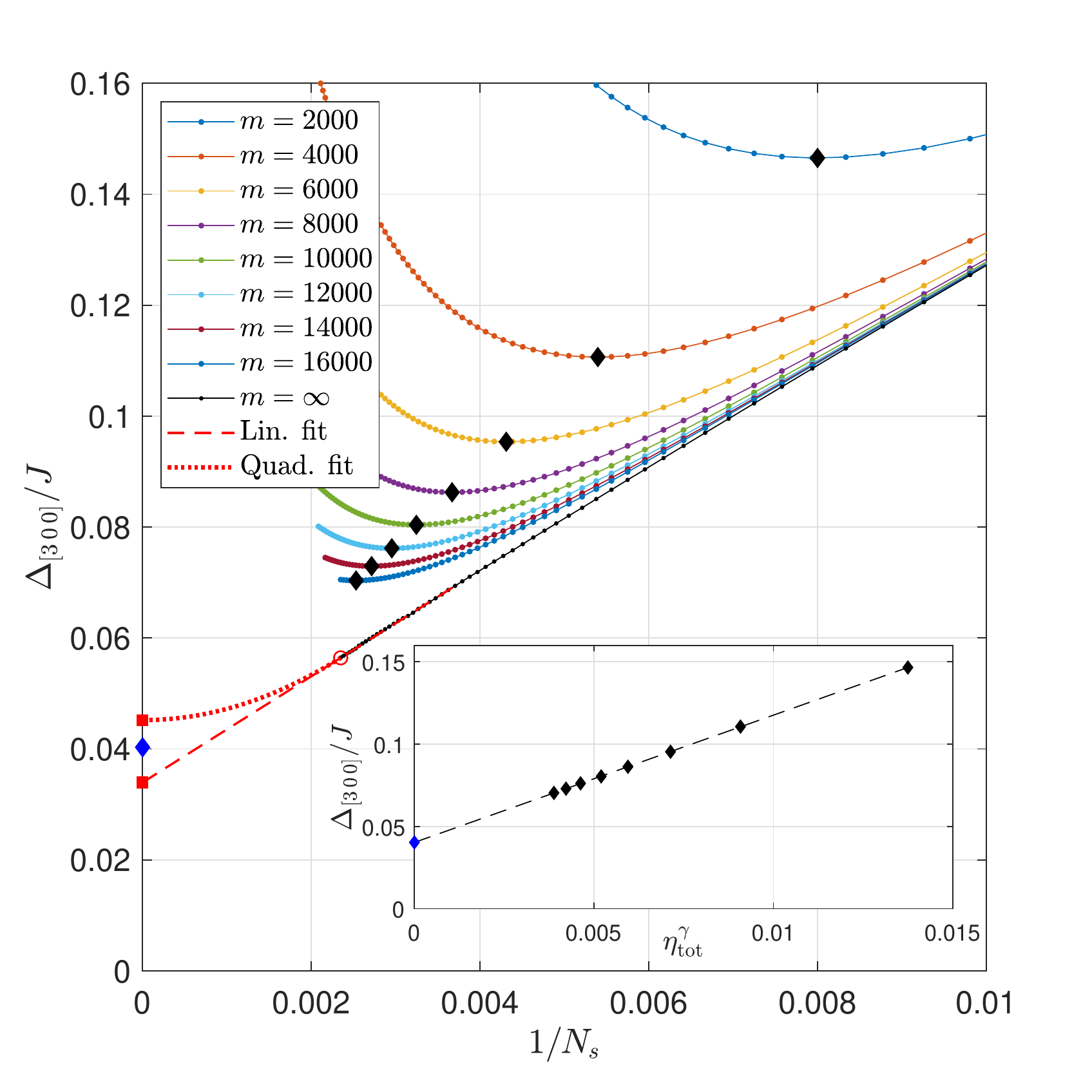}
\end{center}
\caption{Gap from the singlet sector to the symmetric $\irrepsymm$ sector versus inverse chain length for different values of the number $m$ of states kept. The inset shows a power-law fit with exponent $\gamma = 0.47$ of the minimum of each finite-$m$ curve, denoted by a black diamond in the main plot, as a function of the total discarded weight $\eta_{\rm tot}$, which is dominated by the discarded weight in the $\irrepsymm$ sector~\cite{supp_mat}. The extrapolated value, shown with a blue diamond, falls approximately in the middle between our lower and upper bounds of the gap, denoted by red squares.}
\label{fig::gap}
\end{figure}

In our DMRG algorithm, the parameter which mainly controls the accuracy is $m$, the total number of SYTs kept at each step. Each SYT represents a class of wave functions living in the Hilbert space of the half-chain, with the same properties under the action of permutations, but with different $\SU{N}$ weights, so that the color degrees of freedom are factorized out by the use of SYTs \cite{nataf_exact_2014, nataf_density_2018,supp_mat}.
The complexity of our algorithm is then dictated by the diagonalization of the superblock Hamiltonian of dimension $m^2$. In this work we take $m$ to be as large as $m=16\,000$: the discarded weight is then less than $10^{-7}$ in the singlet sector and less than $10^{-5}$ in the $\irrepsymm$ sector~\cite{supp_mat}. 
This gives an accuracy equivalent to the one obtained with over $860\,000$ states in a code which does not keep track of the $\SU{N}$ symmetry. Our main results are summarized in Fig.~\ref{fig::gap}.

Two strategies have been used to extract information about the gap. The first one follows closely the paper of Schollw\"ock {\it et al.} about the spin-$2$ chain~\cite{schollwock_s2antiferro_1996}. For a given number of sites, the energies in each sector are extrapolated as a function of the discarded weight~\cite{supp_mat}. The resulting gap curve remains linear for the largest system sizes, which shows that the size beyond which this curve should bend if there is a gap has not been reached yet. However, and most importantly, this linear curve extrapolates to a finite value. If the system was critical, this curve should extrapolate to zero. In fact, as noted by Schollw\"ock {\it et al.}, the extrapolated value is a lower bound of the gap. Qualitatively, this is the most important result of the present Letter: the spectrum of the three-box symmetric $\SU{3}$ chain is gapped. To get an upper bound, the curvature is assumed to develop for systems immediately larger than the largest system for which we could extrapolate the finite-$m$ results, and we plot a parabola tangent to the gap curve for the largest system size with zero slope in the limit of $1/N_s \rightarrow 0$. The intersection of this parabola with the vertical axis is the upper bound. This analysis leads to the estimate $\Delta_{\irrepsymm}/J \in [0.034,0.046]$.

The other strategy is inspired by the investigation of the Haldane gap in spin-$1$ and spin-$2$ chains by Tatsuaki~\cite{tatsuaki_interaction_2000}: for a fixed $m$, the gap goes through a minimum as a function of the size. The value at the minimum is an estimate for the gap for a given $m$, and this estimate can be extrapolated as a function of the discarded weight, see inset of Fig.~\ref{fig::gap}. The results for $\SU{3}$ appear to follow very accurately a power law with exponent $\gamma = 0.47$, and the extrapolated value $\Delta_{\irrepsymm}/J \simeq 0.040$ is in good agreement with the above bounds.

Note that since the discarded weight stops increasing for a given $m$ for sufficiently long chains, one can also extract the ground state energy per site in the thermodynamic limit by extrapolating the saturated energy per added bond with respect to the saturated discarded weight~\cite{white_numerical_1993}. We obtain $\epsilon/J = -2.176\,397\,3(2)$~\cite{supp_mat}.

\begin{figure}
\begin{center}
\includegraphics[width=0.5\textwidth]{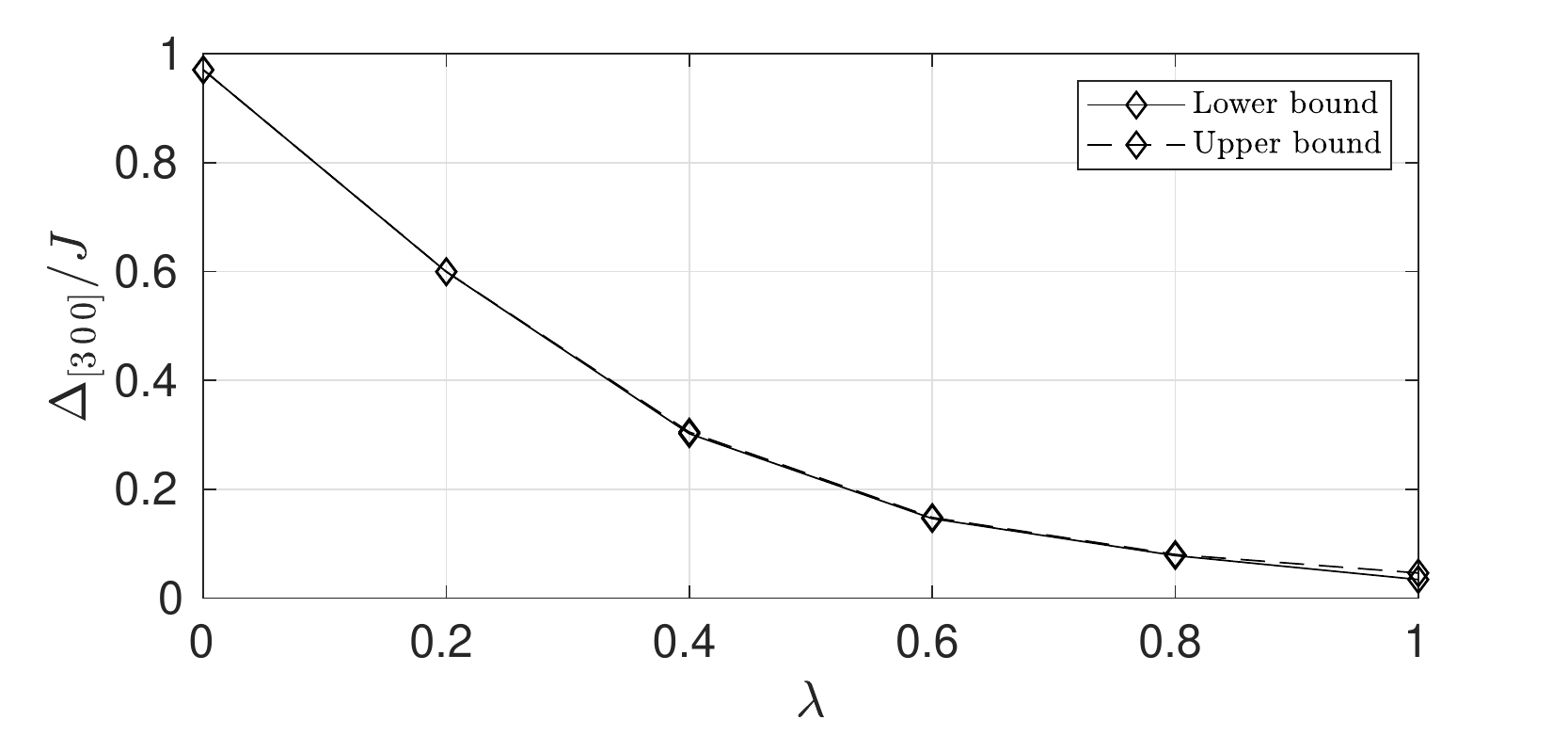}
\end{center}
\caption{Gap from the singlet sector to the symmetric $\irrepsymm$ sector of the interpolation Hamiltonian $\mathcal{H}_{\lambda} = (1-\lambda) \mathcal{H}_{\rm AKLT} + \lambda \mathcal{H}, \ \lambda\in [0,1]$.}
\label{fig::interpol_gap}
\end{figure}

As a further check of the existence of a finite gap in the Heisenberg chain, we have studied the evolution of the gap between the AKLT point, Eq.~\eqref{eq:su3_H_aklt}, and the Heisenberg point Eq.~\eqref{eq:su3-hw-H}. At the AKLT point the correlation length 
is given by $\xi=1/\ln 5 \simeq 0.62$~\cite{gozel_novel_2019}. It is very short, and accordingly the gap is expected to be quite large. Indeed using the same analysis as in Fig.~\ref{fig::gap} we extract the AKLT gap, $\Delta_{\irrepsymm}/J \in [0.970\,705, 0.970\,719]$ using no more than $m=2000$ states~\cite{supp_mat}. Away from the AKLT point the gap decreases smoothly to the value we found at the Heisenberg point, as shown in Fig.~\ref{fig::interpol_gap}.

\begin{figure}
\begin{center}
\includegraphics[width=0.5\textwidth]{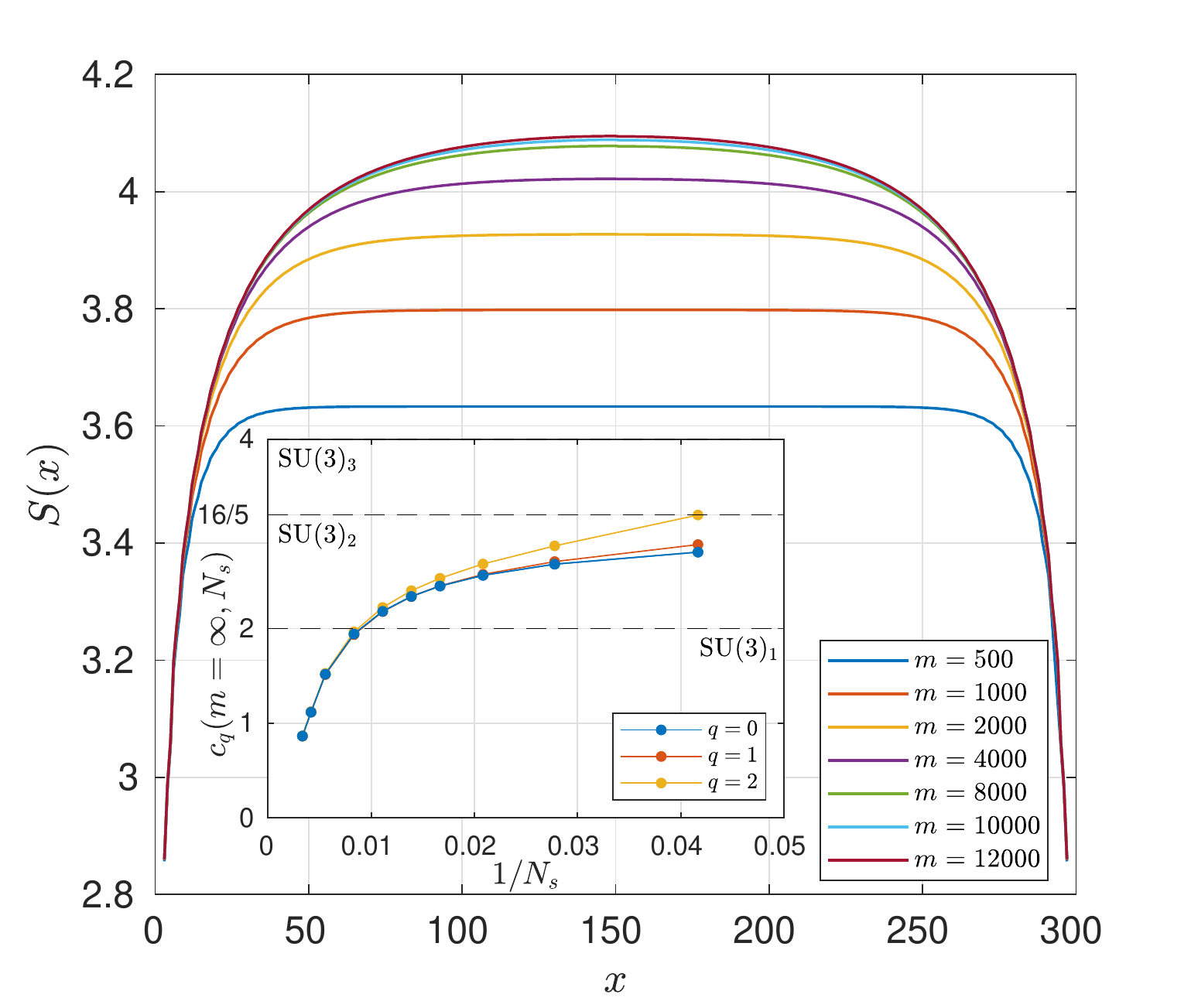}
\end{center}
\caption{Entanglement entropy of the ground state for a chain with $N_s = 300$ sites. Over $839\,000$ states would be needed to reproduce the curve $m=12\,000$ with a code which does not have the $\SU{3}$ symmetry. The inset shows the scaling of the extrapolated central charge in the middle of the chain for different values of the chain length, and comparison with the central charges for the $\SU{3}_k$ WZW conformal field theory with $k=1,2,3$. There are three sets of data $c_q(m=\infty,N_s), \ q=0,1,2$ because of the oscillation of the entanglement entropy along the chain for $x=q  \mod 3$ where $x$ is the position of the cut.}
\label{fig::EE_Ns300_central_charge}
\end{figure}

In view of the conflicting results between the early DMRG results~\cite{rachel_spinon_2009} and the ED results on systems up to $12$ sites~\cite{nataf_exact_2016}, 
we have investigated the entanglement entropy, and we have extracted the central charge using the Calabrese-Cardy formula~\cite{calabrese_entanglement_2004}. For systems as large as $300$ sites, the entanglement entropy still has a significant curvature for $m$ large enough, as can be seen in Fig.~\ref{fig::EE_Ns300_central_charge}, and the finite size estimate of the central charge is not negligible, but it is clearly below the value $c=2$ for the WZW $\SU{3}_1$ universality class, the only alternative to a gapped spectrum~\cite{lecheminant_massless_2015}. Moreover, the results are consistent with a vanishing value in the thermodynamic limit, as expected for a gapped system.

So there is ample evidence that there is a gap in the $\irrepsymm$ sector of the three-box symmetric $\SU{3}$ chain. Let us now discuss the implications of this result for the low-energy spectrum of the model. This discussion relies on two propositions: (i) There are five branches of bulk elementary excitations belonging to four irreps, including $\irrepsymm$; (ii) The presence of a finite gap in any of these irreps, and in particular $\irrepsymm$, implies that there is also one in all other irreps, and hence the spectrum is gapped.

(i) The presence of five branches of bulk elementary excitations is best understood by looking at the ground state of the AKLT model, which has singlets on every bond and which is pictured in Fig.~\ref{fig::aklt_state_su3}(a). To create a bulk excitation one needs to break one singlet, liberating an adjoint irrep on each side of the broken singlet. These adjoint representations can then recombine according to Fig.~\ref{fig::aklt_state_su3}(b) to form five excited states belonging to the $\irrepadj$ (two states), $\irrepsymm$, $\irrepconj$ and $[4\,2\,0]$ irreps~\footnote{This is to be contrasted to the spin-$1$ chain in which the two spin-$1/2$s liberated by breaking a singlet can only recombine into a spin-$1$, leading to a unique spin-$1$ magnon state.}. 

(ii) The presence of a gap in a sector $\alpha$ implies that there is also a gap in another sector $\beta$ if, by combining two or more $\beta$ excitations, one can construct a state that belongs to the $\alpha$ sector. 
Now, the symmetric irrep $\irrepsymm$ appears in the product of two adjoint irreps, as well as in that of two $\irrepconj$ or two $[4\,2\,0]$ irreps. Thus the presence of a gap in the $\irrepsymm$ sector implies that there is a gap in all other sectors of elementary excitations,
in agreement with additional results we have obtained for the $\irrepconj$ irrep~\cite{supp_mat}, hence that the spectrum is gapped.
 This discussion implies that there are actually four Haldane gaps $\Delta_{\alpha}$ corresponding to the four irreps $\alpha = \irrepadj, \irrepsymm, \irrepconj, [4\,2\,0]$, and that they must satisfy 
$\Delta_{\irrepadj},\Delta_{\irrepconj},\Delta_{[4\,2\,0]}\ge \Delta_{\irrepsymm}/2$.
Note that 
any of the irreps of the elementary excitations can be obtained by combining two or more excitations of the other irreps, leading to other inequalities, and the presence of a gap in any of these irreps is a necessary and sufficient condition for a gapped spectrum.

We can actually prove that these inequalities are strict in the case of the AKLT model in Eq.~\eqref{eq:su3_H_aklt} because the lowest excitation in the sector $\irrepsymm$ is not a composite one. Indeed, if it were, the bond energy on a finite chain should show a double peak structure, as observed in the spin-$2$ sector of the spin-$1$ chain~\cite{sorensen_large_1993,supp_mat}, and it does not, as clearly demonstrated by DMRG results on a $60$-site chain~\cite{supp_mat}.

To summarize, using finite-chain DMRG simulations with full $\SU{N}$ symmetry, we have obtained clear numerical evidence that the spectrum of the three-box symmetric $\SU{3}$ chain is gapped, in agreement with field theory arguments, and we have estimated the gap in the $\irrepsymm$ sector to be $\Delta_{\irrepsymm}/J=0.040\pm0.006$. The smallest gap is at least half this gap, hence bounded from below   by $0.017 J$ (half the lower bound of $\Delta_{\irrepsymm}$), and at most $0.046 J$ (the upper bound of $\Delta_{\irrepsymm}$). These bounds point to a very large correlation length of a few hundred sites.

Finally, let us comment on the possible experimental implementation of this model.
Fermions with an $\SU{3}$ degree of freedom can be obtained with $^{87}{\rm Sr}$ or $^{173}{\rm Yb}$ atoms after selecting three out of the ten respectively six nuclear states~\cite{desalvo_degenerate_2010,
taie_realization_2010,
taie_su6_2012}, and protocols are well documented to implement irreps with up to two columns~\cite{gorshkov_two_2010}. Building on Hund's rule that allows one to realize large spins with spin-$1/2$ electrons, a possible route to implement a symmetric irrep with three boxes could be to create a Mott insulating phase by loading three fermions in different orbitals of the same anharmonic trap since the contact interactions between the fermions is expected to lead to a ground state which is antisymmetric in orbital degrees of freedom and symmetric in color $\SU{3}$ space~\cite{beverland_realizing_2016}. 

\begin{acknowledgments}
We thank Ian Affleck and Steve White for very useful discussions, as well as Mikl\'os Lajk\'o and Kyle Wamer for their critical reading of the manuscript. This work has been supported by the Swiss National Science Foundation.
The calculations have been performed using the facilities of the Scientific IT and Application Support Center of EPFL.
\end{acknowledgments}

\nocite{*}
\bibliographystyle{apsrev4-1}
%

\end{document}


\title{Supplemental Material \\ Haldane Gap of the Three-Box symmetric $\SU{3}$ Chain}
\author{Samuel Gozel}
\affiliation{Institute of Physics, Ecole Polytechnique F\'ed\'erale de Lausanne (EPFL), CH-1015 Lausanne, Switzerland}
\author{Pierre Nataf}
\affiliation{Laboratoire de Physique et Mod\'elisation des Milieux Condens\'es, Universit\'e Grenoble Alpes and CNRS, 25 avenue des Martyrs, 38042 Grenoble, France}
\author{Fr\'ed\'eric Mila}
\affiliation{Institute of Physics, Ecole Polytechnique F\'ed\'erale de Lausanne (EPFL), CH-1015 Lausanne, Switzerland}
\date{\today}

\maketitle

\makeatletter
\renewcommand{\theequation}{S\arabic{equation}}
\renewcommand{\thefigure}{S\arabic{figure}}
\renewcommand{\bibnumfmt}[1]{[S#1]}
\renewcommand{\citenumfont}[1]{S#1}

Section~\ref{sec::Heis_H} of this Supplemental Material describes how to rewrite the Heisenberg Hamiltonian in Eq.~(1) in terms of permutations, and how we incorporate the edge spins. We then rewrite the AKLT Hamiltonian in terms of permutations in Section~\ref{sec::AKLT_H}. Section~\ref{sec::DMRG_alg} is devoted to the density matrix renormalization group (DMRG) algorithm with full $\SU{3}$ symmetry and, in particular, to the calculation of the reduced matrix elements of the interaction between the left and right blocks using the subduction coefficients of the symmetric group. In Section~\ref{sec::extrapol} we explain our extrapolation procedure and illustrate it in the case of the $\irrepsymm$ gap. The effect of the edge coupling is studied in Section~\ref{sec::Jend} while the analysis of the entanglement entropy and the extraction of the central charge are presented in Section~\ref{sec::EE}. In Section~\ref{sec::330_heisenberg} we show that the $\irrepconj$ sector is also gapped by proceeding exactly in the same way as described in the main text for the $\irrepsymm$ irrep. In Section~\ref{sec::res_AKLT} we provide further results at the AKLT point, in particular the analysis of the bond energy along the chain, and a discussion of the low-lying excited states. Finally Section~\ref{sec::su2} contains results for the $\SU{2}$ spin-$1$ chain for direct comparison.

\section{Heisenberg Hamiltonian}
\label{sec::Heis_H}

Let $\sigma$ be an irreducible representation (irrep) of the Lie algebra $\mathfrak{su}(N)$. The dimension $\dim(\sigma)$ of the irrep $\sigma$ can be obtained from the Young diagram of the irrep as
\be
\dim(\sigma) = \prod_{i=1}^p \frac{N+\gamma_i}{l_i}
\ee
where $p$ is the total number of boxes in the Young diagram of the irrep, $\gamma_i$ is the algebraic distance from the main diagonal to the $i$-th box, counted positively (respectively negatively) when going to the right (respectively downwards) and $l_i$ is the hook of the $i$-th box of the Young diagram defined as the number of boxes in the same row on the right plus the number of boxes in the same column below plus one for the box itself.

The Heisenberg interaction between two sites $i$ and $j$ carrying the irrep $\sigma$ takes the form~\footnote{The discussion generalizes straightforwardly to the case where the two irreps at site $i$ and $j$ are different.},
\be
\mathcal{H}_{(i,j)} = 2 \, \bold{T}_i \cdot \bold{T}_{j} \equiv 2 \sum_{a=1}^{N^2-1} T^a_i T^a_j
\label{equ::supp::Hij_0}
\ee
where $T^{a} = D^{\sigma}(t^a), \ a=1,...,N^2-1$ is a $\text{dim}(\sigma)$-dimensional matrix representation of the generator $t^a$ of the Lie algebra $\mathfrak{su}(N)$, and $t^a, \ a=1,...,N^2-1$ are the generators in the defining, or fundamental, irrep. The reason for the factor $2$ in Eq.~\eqref{equ::supp::Hij_0} will be elucidated below. We focus now on the case of $N=3$. A convenient choice of generators $t^a$ in the fundamental irrep of $\mathfrak{su}(3)$ is given by the Gell-Mann matrices,
\be
t^a = \frac{1}{2} \lambda^a, \quad a=1,...,8
\ee
and
\bes
\lambda^1 = 
\begin{pmatrix}
0 & 1 & 0 \\
1 & 0 & 0 \\
0 & 0 & 0
\end{pmatrix}, \quad
\lambda^2 = 
\begin{pmatrix}
0 & -i & 0 \\
i & 0 & 0 \\
0 & 0 & 0
\end{pmatrix}, \quad
\lambda^3 = 
\begin{pmatrix}
1 & 0 & 0 \\
0 & -1 & 0 \\
0 & 0 & 0
\end{pmatrix},
\ees
\bes
\lambda^4 = 
\begin{pmatrix}
0 & 0 & 1 \\
0 & 0 & 0 \\
1 & 0 & 0
\end{pmatrix}, \quad
\lambda^5 = 
\begin{pmatrix}
0 & 0 & -i \\
0 & 0 & 0 \\
i & 0 & 0
\end{pmatrix}, \quad
\lambda^6 = 
\begin{pmatrix}
0 & 0 & 0 \\
0 & 0 & 1 \\
0 & 1 & 0
\end{pmatrix},
\ees
\be
\lambda^7 = 
\begin{pmatrix}
0 & 0 & 0 \\
0 & 0 & -i \\
0 & i & 0
\end{pmatrix}, \quad
\lambda^8 = 
\frac{1}{\sqrt{3}} \begin{pmatrix}
1 & 0 & 0 \\
0 & 1 & 0 \\
0 & 0 & -2
\end{pmatrix}.
\label{equ::sm::gell_mann}
\ee
The generators satisfy the $\mathfrak{su}(3)$ commutation relations
\be
[t^a,t^b] = i f^{abc} t^c
\ee
where $f^{abc}$ are the totally antisymmetric structure constants which can easily be calculated using
\be
f^{abc} = \frac{1}{4i} \text{Tr}\left( [\lambda^a, \lambda^b] \lambda^c \right).
\ee
The generators $t^a$ are thus simply the generalizations of the usual $\mathfrak{su}(2)$ generators in the fundamental spin-$1/2$ representation, $\sigma^x/2$, $\sigma^y/2$ and $\sigma^z/2$ where $\sigma^x, \sigma^y, \sigma^z$ are the Pauli matrices.


When $\sigma = \irrepsymm$ is the $3$-box symmetric irrep of $\mathfrak{su}(3)$ corresponding to a Young diagram made of a single row with $3$ boxes, the matrices $T^a, \ a=1,...,8$ are the generators of the Lie algebra in this $10$-dimensional representation. As such, $T^a$ are $10\times 10$ hermitian traceless matrices satisfying the same algebra as the generators in the fundamental representation, and we shall take the following usual normalization condition~\cite{cornwell_grouptheory},
\be
\text{Tr}\left( T^a T^b \right) = \frac{15}{2} \delta^{a b}.
\label{equ::supp::norma_traceTaTb}
\ee
From the generators $T^a$ one can build $3$ sets of raising and lowering operators,
\be
U^{\pm} = T^1 \pm i T^2, \quad V^{\pm} = T^4 \pm i T^5, \quad W^{\pm} = T^6 \pm i T^7
\ee
as well as two Cartan generators~\footnote{The number of Cartan generators corresponds to the rank of the Lie algebra $\mathfrak{su}(N)$ which is $N-1$.},
\be
H_1 = T^3, \quad H_2 = \frac{2}{\sqrt{3}} T^8.
\label{equ::supp::def_cartan_gen}
\ee
Notice that $H_2$ is here the $10$-dimensional matrix representation of the so-called hypercharge operator known in high-energy physics. We now define the operators $\mathcal{S}^{\alpha\beta}, \ \alpha,\beta=1,2,3$ as follows
\be
\mathcal{S}^{12} = U^+, \quad \mathcal{S}^{13} = V^+, \quad \mathcal{S}^{23} = W^+
\ee
and $\mathcal{S}^{\beta\alpha} = (\mathcal{S}^{\alpha\beta})^{\dagger}$ where the Greek indices $\alpha,\beta=1,2,3$ are the {\it color} indices. The diagonal elements $\mathcal{S}^{\alpha\alpha}, \ \alpha=1,2,3$ are chosen in order to satisfy the following relations
\be
H_a = \sum_{\alpha=1}^3 (h_a)_{\alpha,\alpha} \mathcal{S}^{\alpha\alpha}, \quad a=1,2
\ee
where $h_a, \ a=1,2$ are the Cartan generators in the fundamental irrep, as well as the overall tracelessness condition
\be
\sum_{\alpha=1}^3 \mathcal{S}^{\alpha\alpha} = 0
\label{equ::supp::tracelessness_cond}
\ee
which ensures that we end up with $N^2-1=8$ independent generators in total. We obtain,
\be
\mathcal{S}^{11} = H_1 + \frac{1}{2} H_2 = T^3 + \frac{1}{\sqrt{3}} T^8,
\label{equ::supp::S11}
\ee
\be
\mathcal{S}^{22} = - H_1 + \frac{1}{2} H_2 = - T^3 + \frac{1}{\sqrt{3}} T^8
\label{equ::supp::S22}
\ee
and
\be
\mathcal{S}^{33} = - H_2 = - \frac{2}{\sqrt{3}} T^8.
\label{equ::supp::S33}
\ee
It is now a trivial task to show that
\be
\mathcal{S}^{\alpha\beta}_i \mathcal{S}^{\beta\alpha}_j = \text{Tr}\left(\mathcal{S}_i \mathcal{S}_j \right) = 2 \, \bold{T}_i \cdot \bold{T}_j
\label{equ::supp::SS_to_TT}
\ee
where we have reintroduced the site indices $i,j$ and where a summation over the repeated color indices $\alpha, \beta$ is implicit in the first expression. Moreover one can verify that the generators $\mathcal{S}^{\alpha\beta}$ satisfy the so-called $\SU{N}$ commutation relations
\be
\left[ \mathcal{S}^{\alpha\beta}, \mathcal{S}^{\mu\nu} \right] = \delta^{\mu\beta} \mathcal{S}^{\alpha\nu} - \delta^{\alpha\nu} \mathcal{S}^{\mu\beta}.
\label{equ::supp::sun_commut_relations}
\ee
The Heisenberg interaction between two spins at sites $i$ and $j$ is then rewritten as
\be
\mathcal{H}_{(i,j)} = 2 \, \bold{T}_i \cdot \bold{T}_j = \mathcal{S}^{\alpha\beta}_i \mathcal{S}^{\beta\alpha}_j
\label{equ::supp::T_to_Salphabeta}
\ee
where again summation over repeated color indices is implicit. This expression justifies our initial choice of normalization for the interaction. Notice that although we have taken $N=3$ to illustrate the situation the two previous formulas are absolutely general for any irrep $\sigma$ of $\mathfrak{su}(N)$.\\

We now rewrite the Heisenberg interaction in terms of permutation operators. We shall again consider the general case of $\mathfrak{su}(N)$ with an arbitrary $p$-box local irrep $\sigma$. The generators $\mathcal{S}^{\alpha\beta}$ can simply be rewritten in a natural bosonic representation as
\be
\mathcal{S}^{\alpha\beta} = \sum_{A=1}^p b_{\alpha,A}^{\dagger} b_{\beta,A} - \frac{p}{N} \delta^{\alpha\beta}
\label{equ::supp::Salphabeta_bosons}
\ee
where $[b_{\alpha,A}, b_{\beta,B}^{\dagger}] = \delta_{\alpha\beta} \delta_{AB}$ with the constraint that the total number of bosons is conserved and corresponds to the number of boxes, or particles, in the Young diagram,
\be
\sum_{\alpha} \sum_{A=1}^p b_{\alpha,A}^{\dagger} b_{\alpha,A} = p.
\label{equ::supp::constraint_bosons}
\ee
The $\SU{N}$ commutation relations in Eq.~\eqref{equ::supp::sun_commut_relations} are satisfied by the representation~\eqref{equ::supp::Salphabeta_bosons} where the second term simply preserves the tracelessness condition in Eq.~\eqref{equ::supp::tracelessness_cond} once Eq.~\eqref{equ::supp::constraint_bosons} is taken into account. Let us use $\Gamma_{i}$ to denote the set of all particles living on site $i$ in a global numbering. For instance, for two sites $i=1$ and $j=2$ with the $3$-box symmetric irrep $\sigma=\irrepsymm$ one could chose $\Gamma_1 = \left\{ 1, 2, 3 \right\}$ and $\Gamma_2 = \left\{ 4, 5, 6 \right\}$. The Heisenberg interaction can then simply be rewritten as~\footnote{Again, the more general case of two different local irreps $\sigma_i$ and $\sigma_j$ with $p_i$ and $p_j$ boxes in their Young diagrams, respectively, is straightforwardly obtained.}
\be
\mathcal{H}_{(i,j)} = \sum_{A_i\in \Gamma_i} \sum_{A_j' \in \Gamma_j} \mathcal{P}_{A_i,A_j'} - \frac{p^2}{N}
\label{equ::supp:Hij_avec_cst_0}
\ee
where
\be
\mathcal{P}_{A_i,A_j'} = \sum_{\mu\nu} b_{\mu,A,i}^{\dagger} b_{\nu,A',j}^{\dagger} b_{\nu,A,i} b_{\mu,A',j}.
\ee
The operator $\mathcal{P}_{A_i,A_j'}$ is the {\it permutation operator} which interchanges particle $A_i \in \Gamma_i$ with particle $A_j' \in \Gamma_j$. We have thus shown that, up to a constant term, the Heisenberg interaction is fully equivalent to an operator which interchanges (or permutes) all particles of site $i$ with all particles of site $j$. In what follows we shall remove the constant term in Eq.~\eqref{equ::supp:Hij_avec_cst_0}.\\

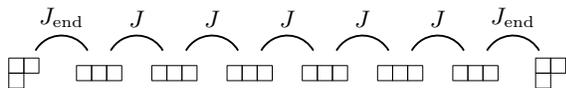
\begin{figure}
\begin{center}
\begin{tikzpicture}
\draw (0.0,0.0) node{$\shsc{0pt}{0.4}{\ydiagram{2,1}}$};
\draw (1.0,0.0) node[]{$\shsc{0pt}{0.4}{\ydiagram{3}}$};
\draw (2.0,0.0) node[]{$\shsc{0pt}{0.4}{\ydiagram{3}}$};
\draw (3.0,0.0) node[]{$\shsc{0pt}{0.4}{\ydiagram{3}}$};
\draw (4.0,0.0) node[]{$\shsc{0pt}{0.4}{\ydiagram{3}}$};
\draw (5.0,0.0) node[]{$\shsc{0pt}{0.4}{\ydiagram{3}}$};
\draw (6.0,0.0) node[]{$\shsc{0pt}{0.4}{\ydiagram{3}}$};
\draw (7.0,0.0) node[]{$\shsc{0pt}{0.4}{\ydiagram{2,1}}$};
\centerarc[black,line width=0.7pt](0.5,0.1)(30:150:0.4)
\draw (0.5,0.5) node[above]{\small $J_{\rm end}$};
\centerarc[black,line width=0.7pt](1.5,0.1)(30:150:0.4)
\draw (1.5,0.5) node[above]{\small $J$};
\centerarc[black,line width=0.7pt](2.5,0.1)(30:150:0.4)
\draw (2.5,0.5) node[above]{\small $J$};
\centerarc[black,line width=0.7pt](3.5,0.1)(30:150:0.4)
\draw (3.5,0.5) node[above]{\small $J$};
\centerarc[black,line width=0.7pt](4.5,0.1)(30:150:0.4)
\draw (4.5,0.5) node[above]{\small $J$};
\centerarc[black,line width=0.7pt](5.5,0.1)(30:150:0.4)
\draw (5.5,0.5) node[above]{\small $J$};
\centerarc[black,line width=0.7pt](6.5,0.1)(30:150:0.4)
\draw (6.5,0.5) node[above]{\small $J_{\rm end}$};
\end{tikzpicture}
\end{center}
\caption{Illustration of the chain with open boundary conditions that we study in this paper.}
\label{fig::supp::1d_chain_with_edge}
\end{figure}

Here we focus on the case of $\SU{3}$ with local $10$-dimensional irrep $\shsc{0pt}{0.4}{\ydiagram{3}}$ at each site. As explained in the main text we screen the edge states by adding edge spins transforming in the $8$-dimensional adjoint representation of $\SU{3}$ at both ends of the open chain, as pictured in Fig.~\ref{fig::supp::1d_chain_with_edge} for $N_s = 8$ sites. The Hamiltonian corresponding to this chain and which we deal with in this paper is thus expressed as,
\be
\mathcal{H} =  J_{\rm end} \left( \mathcal{H}_{(1,2)} + \mathcal{H}_{(N_s-1,N_s)} \right) + J \sum_{i=2}^{N_s-2} \mathcal{H}_{(i,i+1)}
\label{equ::supp:H_Heisenberg_final}
\ee
with positive antiferromagnetic couplings $J,J_{\rm end}>0$ and where
\be
\mathcal{H}_{(i,j)} = \sum_{k\in \Gamma_i} \sum_{l\in \Gamma_j} \mathcal{P}_{k,l}.
\label{equ::supp:Hij_sans_cst}
\ee
Table~\ref{table::supp::ED} provides the lowest energy per site of Hamiltonian~\eqref{equ::supp:H_Heisenberg_final} with $J = J_{\rm end} = 1$ in the sectors with lowest quadratic Casimir obtained by exact diagonalization (ED) on small chains. One observes that the ground state indeed lives, as expected, in the singlet sector while the first excited state is in the adjoint sector.

\begin{table*}
\begin{ruledtabular}
\begin{tabular}{cccccc}
$N_s$ & $[0\,0\,0]$ & $\irrepadj$ & $\irrepsymm$ & $[3\,3\,0]$ & $[4\,2\,0]$ \\
\colrule
4 & -1.974744871391588 & -1.790136694077597 & -1.164271923047740 & -1.648214510756671 & -1.266967328184731 \\
6 & -2.013252996775515 & -1.926290364471545 & -1.785475255588588 & -1.825564289231642 & -1.708066481575317 \\
8 & -2.046690874989769 & -1.986081236571477 & -1.918645070989438 & -1.941440639972563 & -1.881072880540164 \\
10 & -2.071149576814606 & -2.035346364842196 & -1.983620677674574 & -2.000931376373051 & -1.958633675572168 \\
\end{tabular}
\end{ruledtabular}
\caption{Lowest energy per site of Hamiltonian~\eqref{equ::supp:H_Heisenberg_final} with $J = J_{\rm end} = 1$ in the five sectors with lowest quadratic Casimir for chains of length $N_s=4,6,8,10$ obtained by ED. The ground state lives in the singlet sector and the first excited state in the adjoint sector.}
\label{table::supp::ED}
\end{table*}

\section{AKLT Hamiltonian}
\label{sec::AKLT_H}

The $\SU{3}$ AKLT state with $3$-box symmetric irrep at each site was introduced in Ref.~\cite{greiter_valence_2007} building on the projection of $3$ fundamental irreps on each physical site and forming extended singlet bonds on $3$ neighboring sites. 
The parent Hamiltonian of this state was derived in terms of the $10$-dimensional traceless hermitian generators $T^{a}, \, a=1,...,8$ introduced in the previous section as (see Eq.~(55) of Ref.~\cite{greiter_valence_2007})
\be
\mathcal{H}_{\rm AKLT} = 5 J \sum_i \left( \bold{T}_i \cdot \bold{T}_{i+1} + \frac{1}{5} (\bold{T}_i \cdot \bold{T}_{i+1})^2 + \frac{6}{5} \right).
\label{equ::supp::Haklt_greiter}
\ee


In this paper we rely on the construction of the AKLT state introduced in Ref.~\cite{gozel_novel_2019} and illustrated in Fig.~1(a) of the main text. We emphasize that this construction is equivalent to the original construction but has two advantages: i) the nature of the edge states is manifest; ii) it is optimal in the matrix product state sense~\cite{gozel_novel_2019}.

\ytableausetup{nosmalltableaux}

Now we give details on how to rewrite Hamiltonian~\eqref{equ::supp::Haklt_greiter} in terms of permutations in order to use it numerically. We could simply use Eq.~\eqref{equ::supp::T_to_Salphabeta} to replace the $\bold{T}$ generators in Eq.~\eqref{equ::supp::Haklt_greiter} by the $\mathcal{S}^{\alpha\beta}$ operators and then rewrite the Hamiltonian in terms of permutations as was done for the Heisenberg model. However we find it more instructive to rederive the AKLT Hamiltonian in the language of permutations. We do so with the help of the definition of the quadratic Casimir operator in terms of permutations. For any Young diagram $\beta$ the quadratic Casimir is given by~\cite{pilch_formulas_1984},
\be
C_2'(\beta) = \sum_{1\leqslant i < j \leqslant p} \mathcal{P}_{i,j} = \frac{1}{2} \left( \sum_i \beta_i^2 - \sum_j (\beta^T_j)^2 \right)
\label{equ::supp:casimir_permutation}
\ee
where $\mathcal{P}_{i,j}$ is the permutation operator which interchanges particle $i$ with particle $j$, $\beta_i$ is the length of the $i$-th row of the shape $\beta$, and $\beta^T_j$ is the length of the $j$-th column of the shape $\beta$, or, equivalently, the length of the $j$-th row of the transposed shape $\beta^T$, and $p$ is the total number of boxes in the Young diagram. Notice that 
\be
C_2(\beta) = \bold{T} \cdot \bold{T} = C_2'(\beta) + \frac{p}{2} \left( N - \frac{p}{N} \right),
\ee
is the eigenvalue of the quadratic Casimir when the generators $\bold{T}$ are normalized as in Eq.~\eqref{equ::supp::Haklt_greiter}.

We proceed now to the construction of the Hamiltonian. On two neighboring sites with irrep $\beta$ we define the following operator, which interchanges all particles of the $2$ neighboring sites,
\be
\mathcal{P} = \sum_{1\leqslant i < j \leqslant 2 p} \mathcal{P}_{i,j} = 2 \, C_2'(\beta) + \mathcal{H}_{(1,2)}
\ee
where $\mathcal{H}_{(1,2)}$ is our two-site interaction which interchanges particles of site $1$ with particles of site $2$ (see Eq.~\eqref{equ::supp:Hij_sans_cst}). 
Taking $\beta = \irrepsymm$ then the AKLT Hamiltonian between site $1$ and site $2$ takes the form
\be
h_{\rm AKLT} \propto \left( \mathcal{P} - 3 \right) \left( \mathcal{P} - 5 \right) = \left( \mathcal{H}_{(1,2)} + 3 \right) \left( \mathcal{H}_{(1,2)} + 1 \right).
\ee
Choosing the overall scale factor to be $1/4$ we finally end up with,
\be
\mathcal{H}_{\rm AKLT} = J \sum_i \left( \mathcal{H}_{(i,i+1)} + \frac{1}{4} \mathcal{H}_{(i,i+1)}^2 + \frac{3}{4} \right), \ J>0,
\ee
where the interaction $\mathcal{H}_{(i,i+1)}$ is given in Eq.~\eqref{equ::supp:Hij_sans_cst}.

As for the pure Heisenberg model we screen the adjoint edge states which exist for open boundary conditions by adding adjoint spins at both ends of the chain. Removing the superficial constant term the Hamiltonian is finally given by,
\be
\begin{aligned}
\mathcal{H}_{\rm AKLT} = & \ J_{\rm end} \left( \mathcal{H}_{(1,2)} + \mathcal{H}_{(N_s-1,N_s)} \right) \\
& + J \sum_{i=2}^{N_s-2} \left( \mathcal{H}_{(i,i+1)} + \frac{1}{4} \mathcal{H}_{(i,i+1)}^2 \right), \ J,J_{\rm end}>0. \\
\label{equ::supp::aklt_H_final}
\end{aligned}
\ee
The ground state of this Hamiltonian, which is in the singlet sector for $J_{\rm end}/J \simeq 1$, has energy
\be
E_0(N_s) = - 3 \left( 2 J_{\rm end} + \frac{J}{4} (N_s-3) \right).
\ee

Finally, we define the interpolation Hamiltonian
\be
\mathcal{H}_{\lambda} = (1-\lambda) \mathcal{H}_{\rm AKLT} + \lambda \mathcal{H}, \quad \lambda\in [0,1]
\ee
where $\mathcal{H}_{\rm AKLT}$ is given in Eq.~\eqref{equ::supp::aklt_H_final} and $\mathcal{H}$ is given in Eq.~\eqref{equ::supp:H_Heisenberg_final}. Developing one obtains,
\be
\begin{aligned}
\mathcal{H}_{\lambda} = & \ J_{\rm end} ( \mathcal{H}_{(1,2)} + \mathcal{H}_{(N_s-1,N_s)} ) \\
& + J \sum_{i=2}^{N_s-2} \left( \mathcal{H}_{(i,i+1)} + \frac{1-\lambda}{4} \mathcal{H}_{(i,i+1)}^2 \right), \quad \lambda\in [0,1].
\end{aligned}
\ee
Thus, the $\lambda$ parameter simply tunes the permutational biquadratic term $\mathcal{H}_{(i,i+1)}^2$ from its value at the AKLT point to $0$ as $\lambda$ evolves from $0$ to $1$. We insist that this interpolation is natural in the language of permutations.

\section{Description of the DMRG algorithm}
\label{sec::DMRG_alg}

\ytableausetup{smalltableaux}

The DMRG algorithm developed in this paper is an extension of the one designed for the fundamental irrep~\cite{nataf_density_2018}. In order to incorporate the adjoint edge spins we start from a half-chain with $N_s/2-1$ symmetric irreps and one adjoint irrep using the technique developed in Ref.~\cite{wan_exact_2017}. Following the original formulation by White, which is suitable for our implementation of the $\SU{N}$ symmetry, the chain is then increased by adding two sites in the middle and integrating them to the left and right blocks~\cite{white_density_1992,white_density_1993}. The Hamiltonian of the new left (respectively right) block with one site more, is relatively easy to get using the concepts and recipes given in  Ref.~\cite{nataf_exact_2016,wan_exact_2017}.
Both the genealogy of irreps, created using the Itzykson-Nauenberg rules~\cite{itzykson_unitarity_1966}, and the truncation procedure, determines which states should be kept at a given step. 
Such a truncation is controlled by two parameters. The first one is the total number $M$ of irreps in which we allow each half-block to have states in. As the Hamiltonian is antiferromagnetic we expect the low energy states to belong to the irreps with the lowest quadratic Casimir. We thus select the $M$ irreps having the smallest Casimir to form this collection. The parameter $M$ is taken to be $M = 299$ for the singlet sector on the full chain, and $M = 155$ for the symmetric sector $\irrepsymm$ on the full chain. Once $M$ is chosen one needs to compute once for all, but separately for the singlet and the $\irrepsymm$ sectors, the {\it reduced matrix elements}, or matrix elements of the interaction term between the left and the right block using the subduction coefficients associated to these $M$ irreps. We will give below an example of the calculation of such a coefficient. The second parameter is the total number $m$ of standard Young tableaus (SYTs) in each half-block. The SYTs play the role of what is called {\it multiplets} in Ref.~\cite{weichelbaum_nonabelian_2012}, or reduced basis in Ref.~\cite{mcculloch2002}. The $m$ states are then distributed among the $M$ irreps such that the total discarded weight, namely the sum of the discarded eigenvalues of the density matrix in each sector, is minimized. The distribution of the $m$ states among the $M$ irreps allows us to verify that $M$ is large enough for a given $m$. With the chosen values of $M$ in each sector we ensure that the accuracy only depends on $m$ because the irreps with the largest Casimir are never selected to host states. 
We give now an example of the calculation of the matrix elements of the interaction term between the left and right block using the subduction coefficients of the symmetric group.
We focus mainly on the additional steps required to go from the fundamental irrep case treated in great details in Ref.~\cite{nataf_density_2018} and the $3$-boxes symmetric irrep case treated here, and we shall use notations and concepts of both Ref.~\cite{nataf_density_2018} and~\cite{nataf_exact_2016}, to demonstrate that:

\begin{equation}
\text{\raisebox{-2.4ex}{$\mathlarger{
\mathlarger{
\mathlarger{
\mathlarger{
\mathlarger{
\mathlarger{\langle}}}}}}$}}
\ytableaushort{\,\,\times,\,\times,\times} 
 \shsc{-6pt}{1}{$\otimes$} 
\ytableaushort{\,\,\times,\,\times,\times} 
\text{\raisebox{-2.4ex}{$\mathlarger{
 \mathlarger{
 \mathlarger{
 \mathlarger{
 \mathlarger{
 \mathlarger{\vert}}}}}}$}}
 \mathcal{P}_{\text{\small $\ytableaushort{\,\,\,\,,\,\,\,\,,\,\,\,\,}$}} 
\text{\raisebox{-2.4ex}{$\mathlarger{
\mathlarger{
\mathlarger{
\mathlarger{
\mathlarger{
\mathlarger{\vert}}}}}}$}}
\shsc{-4pt}{1}{\ytableaushort{\,\,\times\times,\,\times} }
 \,\shsc{-6pt}{1}{$\otimes$} \, 
\shsc{-4pt}{1}{\ytableaushort{\,\,\times\times,\,\times}}
\text{\raisebox{-2.4ex}{$\mathlarger{
\mathlarger{
\mathlarger{
\mathlarger{
\mathlarger{
\mathlarger{\rangle}}}}}}$}} 
\nonumber 
\end{equation}

\begin{equation}
=\frac{2\sqrt{6}}{5}. \label{eq_coeff}
\end{equation}
In the last equation, the two shapes which appear for instance on the right  part (which we can call  the {\it ket}) of the reduced matrix element are $\shsc{3pt}{0.6}{\ydiagram{4,2}}$ meaning that we consider classes of states belonging to the irrep $\shsc{3pt}{0.6}{\ydiagram{4,2}}$ for both the left and the right half-chain. Generally, the shapes appearing in this kind of coefficients should always contain a number of boxes $3 \times q$, with $q$ some positive integer (here $q=2$).
Note that there is no reason for these two shapes to be the same but we decided to show this example which is quite simple. 
The last coefficient is for instance useful in the infinite DMRG part to build the interaction term between the left and the right half-chain in the singlet sector target space $[4\,4\,4]$, i.e. $ \mathcal{H}_{(\frac{N_s}{2},\frac{N_s}{2}+1)} \equiv \mathcal{P}_{[4\,4\,4]}$ when $N_s/2=q+3p$, where $p$ is another integer. The three crosses inside each shape keep track of the genealogy of the class of states we are interested in: in this example, for the part corresponding to the left half-chain of the ket, the states are built from the shape $\shsc{3pt}{0.6}{\ydiagram{2,1}}$ at the previous stage and are such that the integration of the last site containing irrep $\shsc{0pt}{0.6}{\ydiagram{3}}$ gives states in the shape $\shsc{3pt}{0.6}{\ydiagram{4,2}}$.
We use the SYTs and the orthogonal representation of the symmetric group to perform the calculation shown in Eq.~\eqref{eq_coeff}. 
We follow the same first steps as for the fundamental irrep case~\cite{nataf_density_2018}. Let us focus again on the {\it ket}:
for the left part, we first replace the shape and cross symbols by a SYT $S_1$
 of the same shape having its last three numbers located at the positions of the three crosses and ordered in such a way that it is the {\it smallest} according to the last letter order sequence:
\begin{align}
\shsc{2pt}{1}{\ytableaushort{\,\,\times\times,\,\times}} \ \rightarrow \ \shsc{2pt}{1}{\ytableaushort{1245,36}} \ = S_1.
\end{align}  

In fact, the actual ordering of numbers $1,2$ and $3$ does not really matter, while the positions of the numbers $4,5$ and $6$ 
 correspond to the order that we have chosen to select the {\it representative} of the class of equivalence of SYTs which have $4$, $5$ and $6$ put at the three locations of the crosses (see page 4 of Ref.~\cite{nataf_exact_2016}).

For the second shape, we start from the smallest SYT of the full shape after having deleted the crosses and we reindex the numbers following $(1,2, .., 3 \times q) \rightarrow (3\times 2 q, 3\times 2q-1, .., 3 \times q)$ :
\begin{align}
\shsc{2pt}{1}{\ytableaushort{\,\,\times\times,\,\times}} \ \rightarrow \ \shsc{2pt}{1}{\ytableaushort{\,\,\,\,,\,\,}} \ \rightarrow \ \shsc{2pt}{1}{\begin{ytableau}
 12 &11 &10&9 \\8 &7 \end{ytableau}} \ = S_2.
\end{align}  
Then we expand the ``product'' $S_1 \otimes S_2$ on the shape  $[4\,4\,4]$ using the subduction coefficients~\cite{chen_grouptheory}:
\begin{align}
\shsc{2pt}{1}{\ytableaushort{1245,36}} \, \otimes \,
 \shsc{2pt}{1}{\begin{ytableau}
 12 &11 &10&9 \\8 &7 \end{ytableau}} \ \rightarrow \ \shsc{6pt}{1}{\begin{ytableau}
 1 &2&4&5\\3&6&7&8\\9&10 &11 &12 \end{ytableau}}.
\label{eq_expansion}
\end{align}
To perform such an expansion, we had to solve the nullspace of some operator (see step 2 page 8 of Ref.~\cite{nataf_density_2018}).
The dimension of such a nullspace is equal to the outer multiplicity of the singlet irrep in the tensor product of $[4\,2\,0] \otimes [4\,2\,0]$, which is one.
This is the only case we are currently able to treat so far: we have not yet developed the theory to treat multiplicities strictly higher than one,
which explains why we restrict ourselves to the target irreps $[0\,0\,0]$, $\irrepsymm$, $\irrepconj$ and $\irrepsix$.
Note that in the simple example provided in Eq.~\eqref{eq_expansion}, there is only one SYT of shape $[4\,4\,4]$ in the expansion of the product (with subduction coefficient  equal to $1$). This is of course not general, and more generally we obtain at this stage a linear superposition of SYTs of the same shape (corresponding to the target irrep).
Then, analogously to step 3 page 9 of Ref.~\cite{nataf_density_2018}, we perform the sequence of operations which transform $S_2$ into $S_2'$, where the three numbers $3\times q +1, 3\times q +2, 3\times q +3$
are now located at the locations of the three crosses in the initial right shape of the ket (with the smallest rank according to the last letter order sequence):
\begin{align}
\shsc{2pt}{1}{\begin{ytableau}
 12 &11 &10&9 \\8 &7 \end{ytableau}} \ = S_2 \rightarrow \ \shsc{2pt}{1}{\begin{ytableau}
 12 &11 &9&8 \\10 &7 \end{ytableau}} \ = S_2'
\end{align}  
onto the resulting expansion of $S_1 \otimes S_2$ to obtain:
\begin{align}
\begin{ytableau}
 1 &2&4&5\\3&6&7&8\\9&10 &11 &12 \end{ytableau} \ \text{\raisebox{-1.5ex}{$\rightarrow$}} \
 \begin{ytableau}
 1 &2&4&5\\3&6&7&10\\8&9 &11 &12 \end{ytableau}\text{\raisebox{-1.5ex}{.}}
\end{align}
The next step does not exist for the fundamental irrep case treated in Ref.~\cite{nataf_density_2018} and is really specific to the ($3$-box) symmetric irrep case. It consists in applying the local symmetric projector 
(see Eq.~(43) in Ref.~\cite{nataf_exact_2016}) for the two {\it pseudo-sites} $q$ and $q+1$, to obtain the following linear superposition:
\begin{align}
 \begin{ytableau}
 1 &2&4&5\\3&6&7&10\\8&9 &11 &12 \end{ytableau} \ \text{\raisebox{-1.5ex}{$\rightarrow$}} \ 
 \text{\raisebox{-1.5ex}{$\frac{1}{9}$}} \begin{ytableau}
 1 &2&4&5\\3&6&7&10\\8&9 &11 &12 \end{ytableau}
\text{\raisebox{-1.5ex}{$\,+  \frac{\sqrt{2}}{9}$}} \begin{ytableau}
 1 &2&4&5\\3&6&8&10\\7&9 &11 &12 \end{ytableau}
 \text{\raisebox{-1.5ex}{$\,+ \frac{\sqrt{2}}{3\sqrt{3}}$}}\begin{ytableau}
 1 &2&4&5\\3&6&9&10\\7&8 &11 &12 \end{ytableau} \nonumber \\
  \text{\raisebox{-1.5ex}{$\,+ \frac{\sqrt{2}}{9}$}}\begin{ytableau}
 1 &2&4&6\\3&5&7&10\\8&9 &11 &12 \end{ytableau}
  \text{\raisebox{-1.5ex}{$\,+ \frac{2}{9}$}}\begin{ytableau}
 1 &2&4&6\\3&5&8&10\\7&9 &11 &12 \end{ytableau}
  \text{\raisebox{-1.5ex}{$\,+ \frac{2}{3\sqrt{3}}$}} \begin{ytableau}
 1 &2&4&6\\3&5&9&10\\7&8 &11 &12 \end{ytableau} \nonumber \\
  \text{\raisebox{-1.5ex}{$\,+ \frac{\sqrt{2}}{3\sqrt{3}}$}} \begin{ytableau}
 1 &2&5&6\\3&4&7&10\\8&9 &11 &12 \end{ytableau}
  \text{\raisebox{-1.5ex}{$\,+  \frac{2}{3\sqrt{3}}$}} \begin{ytableau}
 1 &2&5&6\\3&4&8&10\\7&9 &11 &12 \end{ytableau}
\text{\raisebox{-1.5ex}{$ \,+ \frac{2}{3}$}}\begin{ytableau}
 1 &2&5&6\\3&4&9&10\\7&8 &11 &12 \end{ytableau}\text{\raisebox{-1.5ex}{.}} \nonumber \\
\end{align}
Note that, by construction, the last superposition is fully symmetric under the permutations which interexchange $4$, $5$ and $6$ and separately
$7$, $8$ and $9$. 

We then apply $ \mathcal{H}_{(q,q+1)}\equiv \mathcal{P}_{6,7} + \mathcal{P}_{6,8} + \mathcal{P}_{6,9} + \mathcal{P}_{5,7} + \mathcal{P}_{5,8} + \mathcal{P}_{5,9} +\mathcal{P}_{4,7} + \mathcal{P}_{4,8} + \mathcal{P}_{4,9}$ on this superposition using the rules of the orthogonal representation of the symmetric group~\cite{nataf_exact_2014,
young_onsubstitutional_1900,
rutherford_substitutional}.
Finally, we take the scalar product with the linear superposition of SYTs of shape $[4\,4\,4]$ obtained for the {\it bra} after having followed the very same steps as for the {\it ket} (except for the application of $\mathcal{H}_{(q,q+1)}$) to obtain Eq.~\eqref{eq_coeff}.

\ytableausetup{nosmalltableaux}

\section{Extrapolations of DMRG data}
\label{sec::extrapol}

\begin{figure}
\begin{center}
\includegraphics[width=0.5\textwidth]{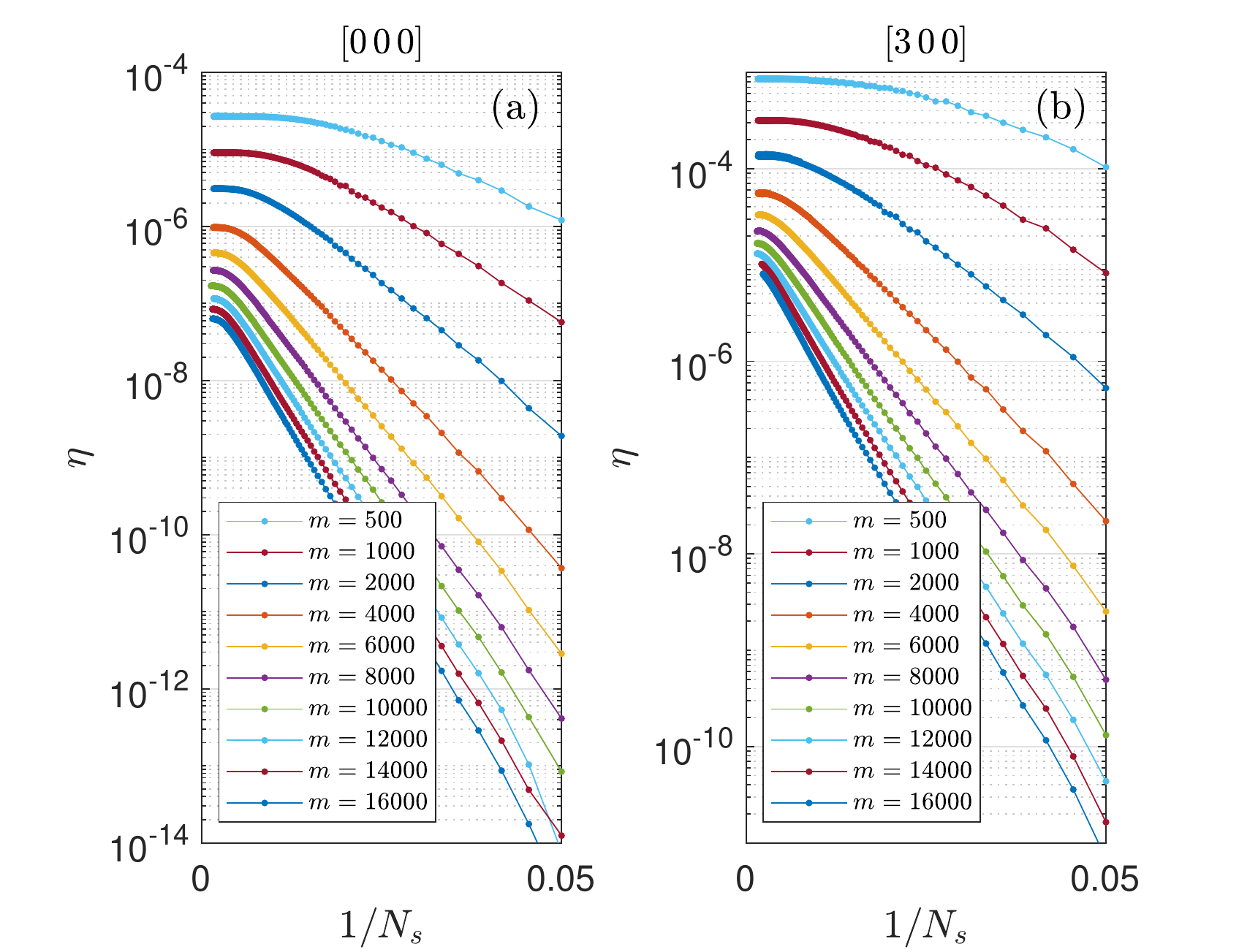}
\end{center}
\caption{Discarded weight versus inverse chain length in (a) the singlet sector and (b) the symmetric sector $\irrepsymm$, for different values of the number of states kept.}
\label{fig::supp::discweight}
\end{figure}

\begin{figure}
\begin{center}
\includegraphics[width=0.5\textwidth]{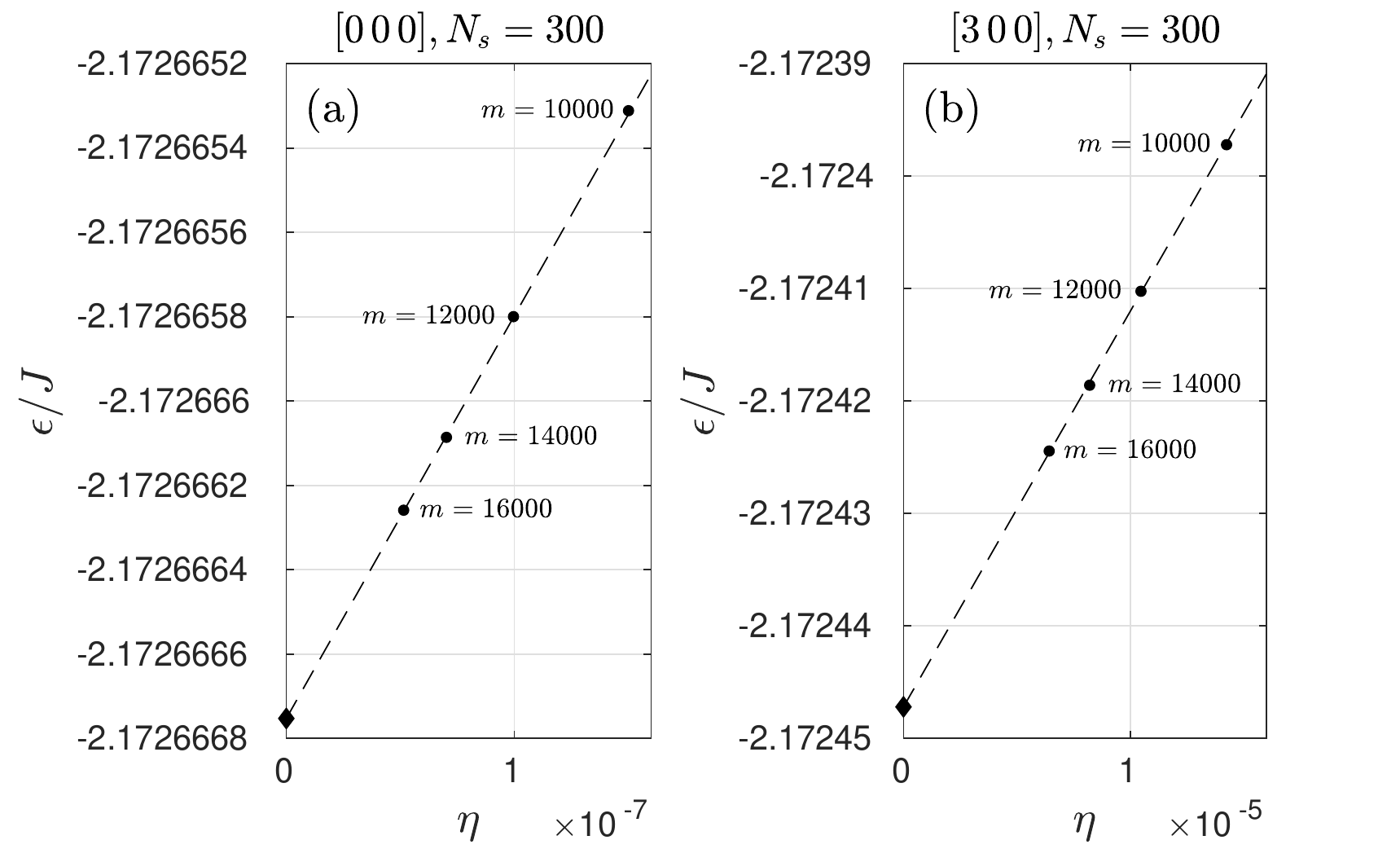}
\end{center}
\caption{Examples of extrapolation of the energy per site versus the  discarded weight at $N_s=300$ for (a) the singlet sector and (b) the symmetric sector $\irrepsymm$. The extrapolated values, denoted by a black diamond, are then used to obtain the gap at $N_s=300$.}
\label{fig::supp::extrapol_discweight}
\end{figure}

The large number of states kept at the Heisenberg point is not sufficient to obtain full convergence of the gap for finite size. Our results thus rely on an extrapolation in the number of states kept, followed by an extrapolation to the thermodynamic limit. Here we discuss both of these extrapolations. The finite-$m$ extrapolation is performed using the discarded weight instead of simply using $1/m$ as a measure of the accuracy of the simulation. Figure~\ref{fig::supp::discweight} shows the discarded weight in the singlet and $\irrepsymm$ sectors at the Heisenberg point. In order to obtain the extrapolated ``$m=\infty$'' gap curve in Fig.~1 of the main text, we do not directly extrapolate the gap but instead we extrapolate separately the lowest energy per site in each sector, as shown in Fig.~\ref{fig::supp::extrapol_discweight}. For the extrapolation to the thermodynamic limit, we follow Schollw\"ock {\it et al.} in Ref.~\cite{schollwock_s2antiferro_1996}. According to standard theory, the gap of a non-critical Hamiltonian should behave quadratically at large distance $N_s \gg \xi/a$ with $\xi = v/\Delta_{\infty}$ the correlation length, $v$ a factor having the dimension of a velocity, $a$ the lattice spacing and $\Delta_{\infty}$ the gap in the thermodynamic limit,
\be
\Delta(N_s) = \Delta_{\infty} + \frac{v^2 \pi^2}{\Delta_{\infty} (N_s a)^2}.
\ee
At shorter distance $N_s \ll \xi/a$ the behavior is linear, $\Delta(N_s) \propto N_s^{-1}$. In our DMRG data in Fig.~1 we clearly see that we have not yet reached the quadratic regime. We thus extrapolate to the thermodynamic limit in two manners. First we extrapolate linearly and obtain a lower bound for the gap. Secondly we assume that the quadratic regime develops just after the largest size that we have reached and fit a parabola tangent to that point with vanishing slope at $1/N_s=0$~\cite{schollwock_s2antiferro_1996}. This provides an upper bound for the gap. These two fits to the thermodynamic limit are shown in Fig.~1 as red dashed and dotted lines, respectively, with their extrapolated values shown as red squares, and the transition point from linear to parabolic regimes is displayed with a red circle. Increasing the number of states kept thus allows one  to decrease the upper bound of the gap, until the quadratic regime is finally observed in the extrapolated gap curve.

Our second extrapolation method for the gap (see inset of Fig. 1) relies on the fact any finite-$m$ gap curve goes through a minimum as a function of the size~\cite{schollwock_s2antiferro_1996,
wang_haldane_1999,
tatsuaki_interaction_2000}. The extrapolation procedure is explained in the main text and we shall only discuss the reason for this minimum.

At finite-$m$, and for sufficiently large systems, there is a competition between the decrease of the gap because of the increase of the system size, and the increase of the accumulated error on the numerical gap. For $m$ fixed and when the system size increases, at some point the accumulated error on the numerical gap value becomes larger than the decrease of the gap because of the increase of the system size. Thus the gap curve saturates and then starts increasing. That is why we use the value of the gap at the minimum as the best possible upper bound of the true gap for finite $m$.

\begin{figure}
\begin{center}
\includegraphics[width=0.5\textwidth]{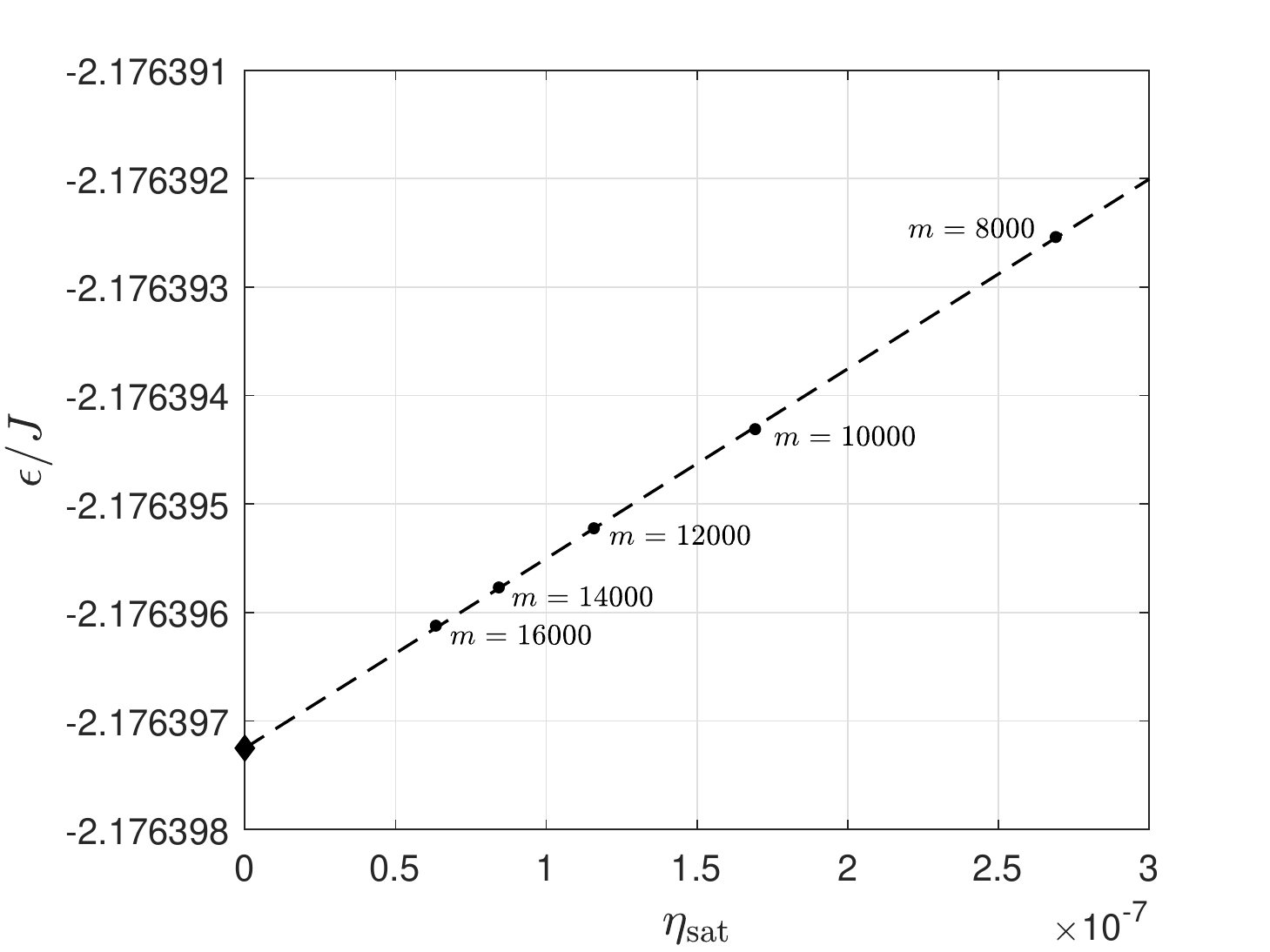}
\end{center}
\caption{Extrapolation of the ground state energy per site in the singlet sector with respect to the saturated discarded weight.}
\label{fig::supp::extrapol_gs_energy_singlet}
\end{figure}

We also extrapolate directly the ground state energy per site by computing the energy of the added bond for each $N_s$~\cite{white_numerical_1993}. This value saturates for a given $m$ once the chain length has reached a given size. Moreover, as can be seen in Fig.~\ref{fig::supp::discweight} the discarded weight also saturates at some point. We thus extrapolate the energy per bond with respect to the saturated discarded weight in Fig.~\ref{fig::supp::extrapol_gs_energy_singlet} and obtain the thermodynamic energy per site in the singlet sector to be $\epsilon/J = -2.1763973(2)$. We notice that the discarded weight saturates for chain lengths well beyond the position of the minimum of the gap. Obtaining Fig.~\ref{fig::supp::extrapol_gs_energy_singlet} thus requires significant additional computing time.

Finally, let us notice that for the symmetric sector $\irrepsymm$ the discarded weight for $m=16000$ is still of order $10^{-5}$, as shown in Fig.~\ref{fig::supp::discweight} and~\ref{fig::supp::extrapol_discweight}, which reveals how difficult it is to tackle this problem numerically. However Fig.~\ref{fig::supp::extrapol_discweight} also shows that the extrapolation with respect to the discarded weight $\eta\rightarrow 0$ is performed accurately.

\section{Effect of the edge coupling on the Heisenberg chain}
\label{sec::Jend}

\begin{figure}
\begin{center}
\includegraphics[width=0.5\textwidth]{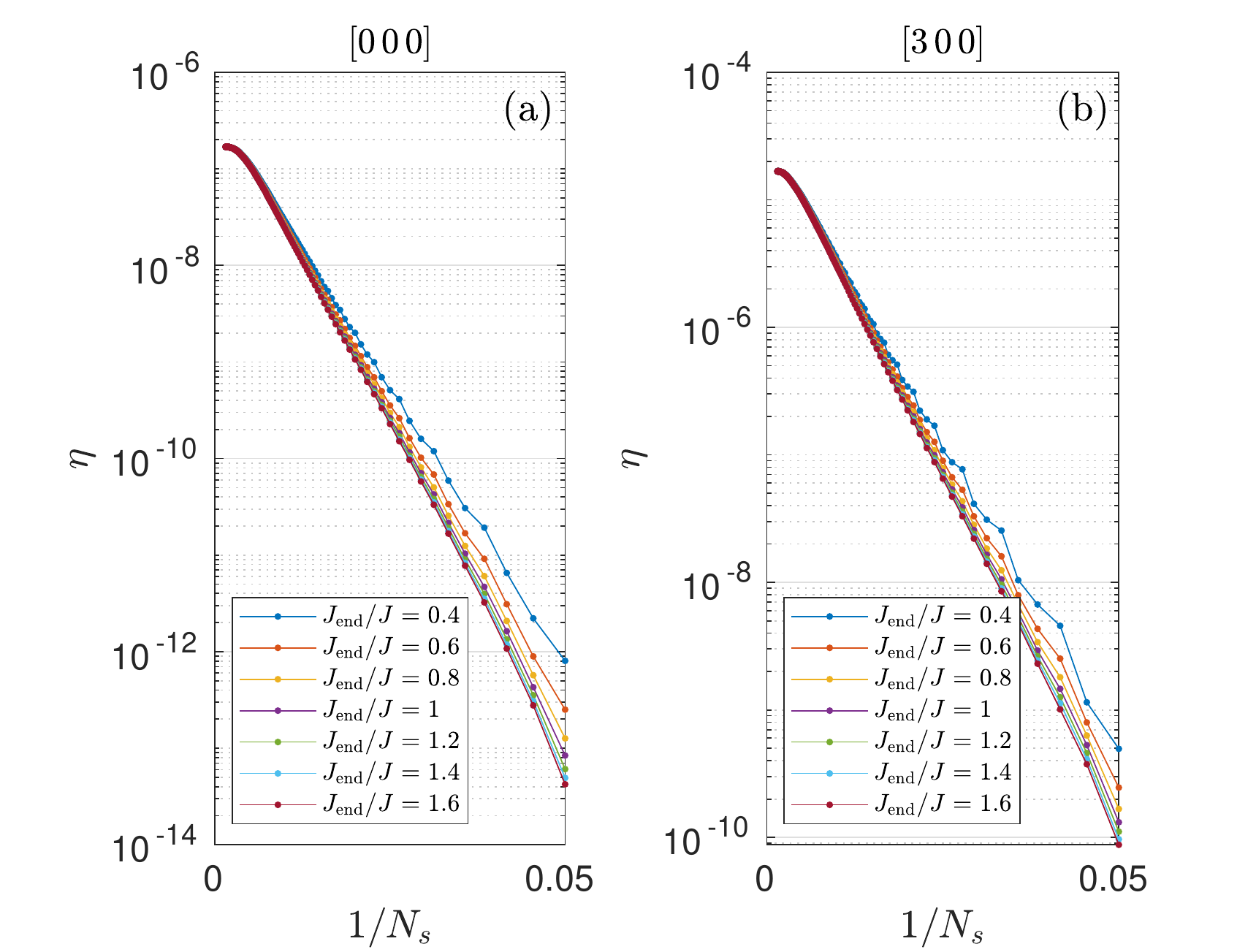}
\end{center}
\caption{Discarded weight versus inverse chain length in (a) the singlet sector and (b) the symmetric sector $\irrepsymm$, for $m=10000$ states kept and different values of the edge coupling $J_{\rm end}$.}
\label{fig::supp::Jend::discweight}
\end{figure}

\begin{figure}
\begin{center}
\includegraphics[width=0.5\textwidth]{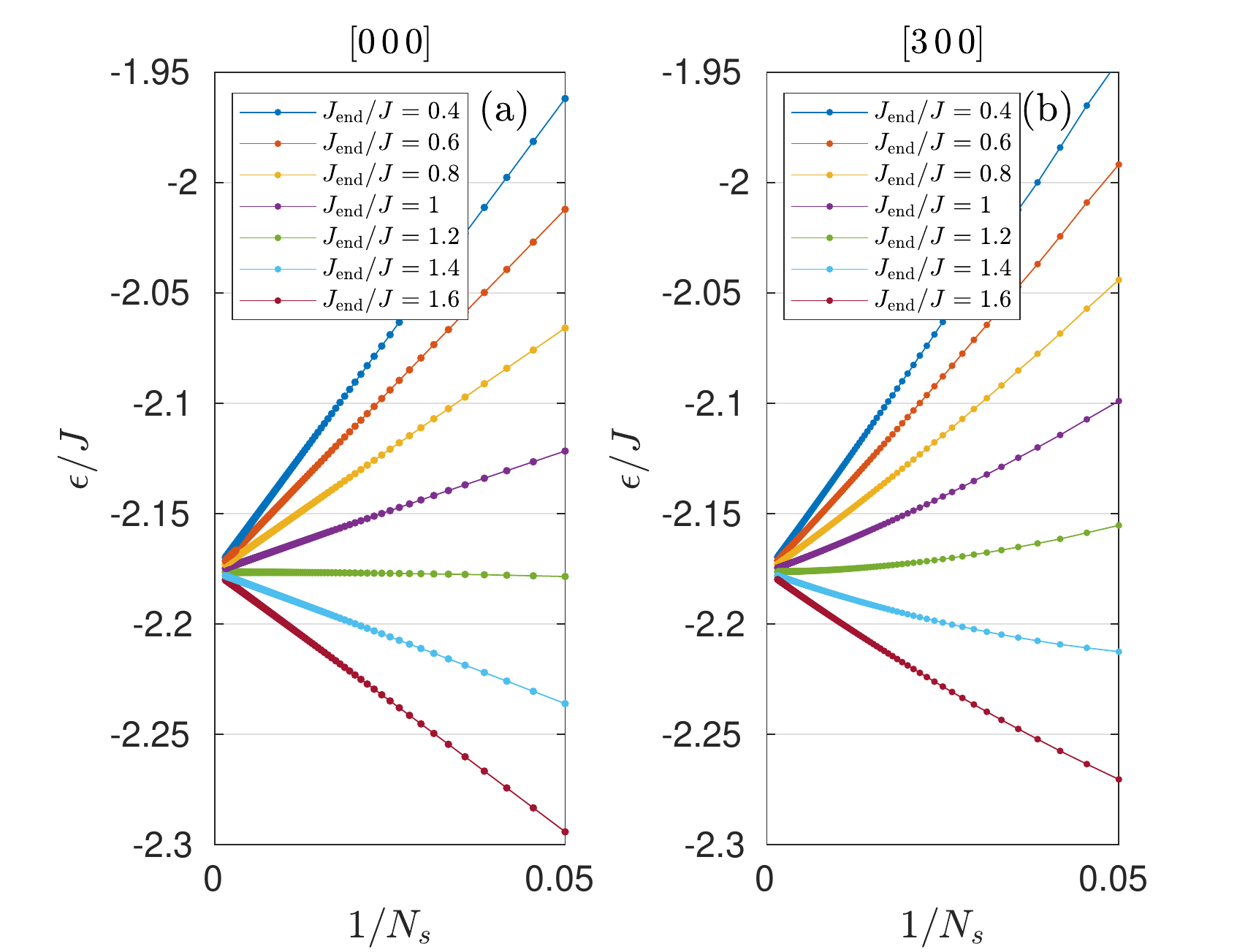}
\end{center}
\caption{Energy per site versus inverse chain length in (a) the singlet sector and (b) the symmetric sector $\irrepsymm$, for $m=10000$ states kept and different values of the edge coupling $J_{\rm end}$.}
\label{fig::supp::Jend::energy}
\end{figure}

\begin{figure}
\begin{center}
\includegraphics[width=0.5\textwidth]{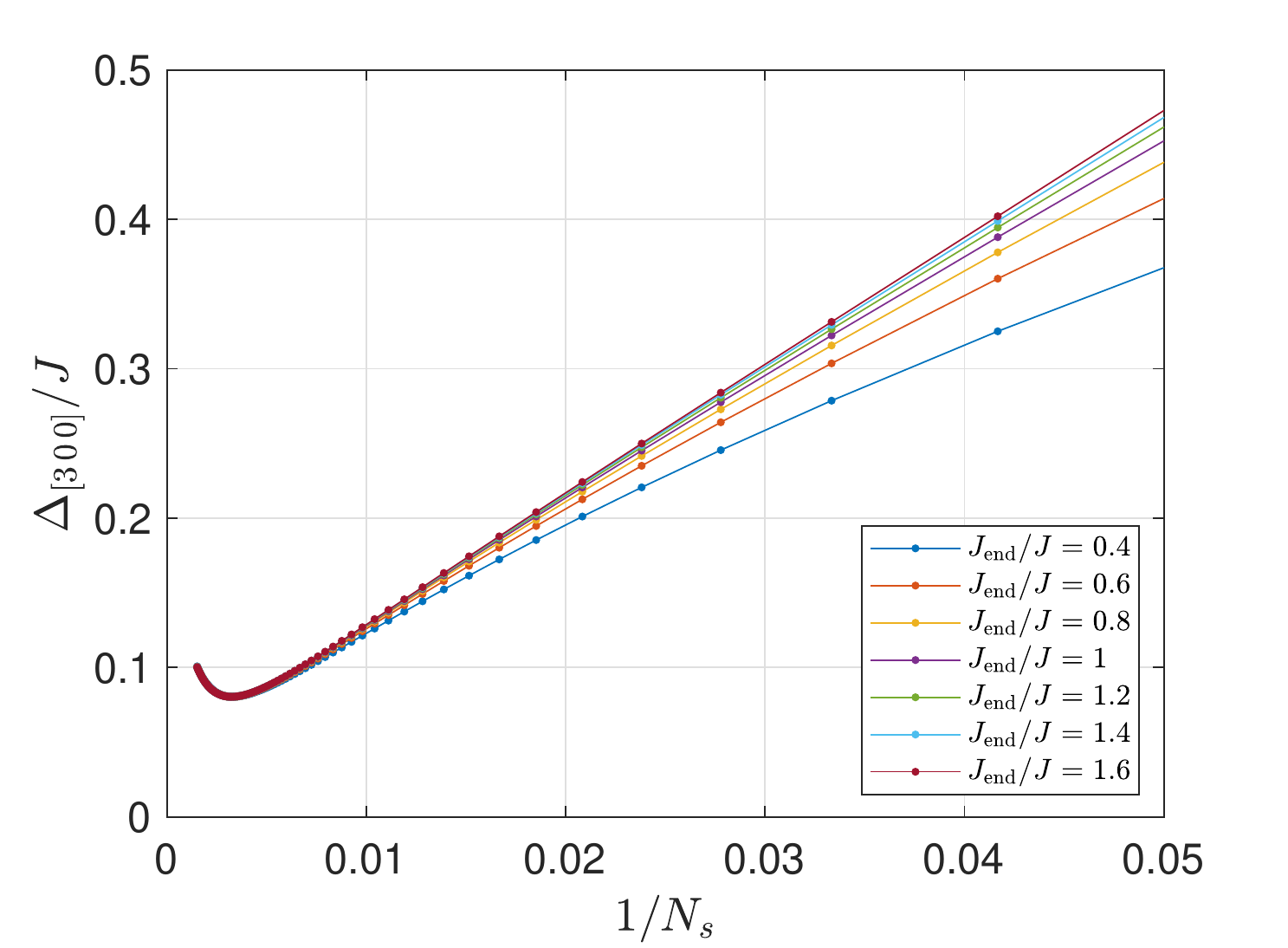}
\end{center}
\caption{Finite-size gap from the singlet sector to the symmetric sector $\irrepsymm$, for $m=10000$ states kept and different values of the edge coupling $J_{\rm end}$.}
\label{fig::supp::Jend::gap}
\end{figure}

We study now the effect of the edge coupling $J_{\rm end}$ in Eq.~\eqref{equ::supp:H_Heisenberg_final}. Although for infinite size the precise value of the edge coupling is inconsequential, for finite size it does affect the lowest energy in each sector and, to a lesser extent, the discarded weight. Figure~\ref{fig::supp::Jend::discweight} obtained with $m=10000$ states kept indeed shows that the discarded weight in the singlet and $\irrepsymm$ sectors are hardly influenced by the edge coupling. The lowest energy per site in each sector is presented in Fig.~\ref{fig::supp::Jend::energy}. From this figure we deduce that  $J_{\rm end}/J \simeq 1.2$ would be an optimal value as the curves are the most flat in each panel. However when looking at the gap given in Fig.~\ref{fig::supp::Jend::gap} one sees that all values of $J_{\rm end}/J$ lead precisely to the same minimum of the gap.

\section{Entanglement entropy and central charge}
\label{sec::EE}

\begin{figure}
\begin{center}
\includegraphics[width=0.5\textwidth]{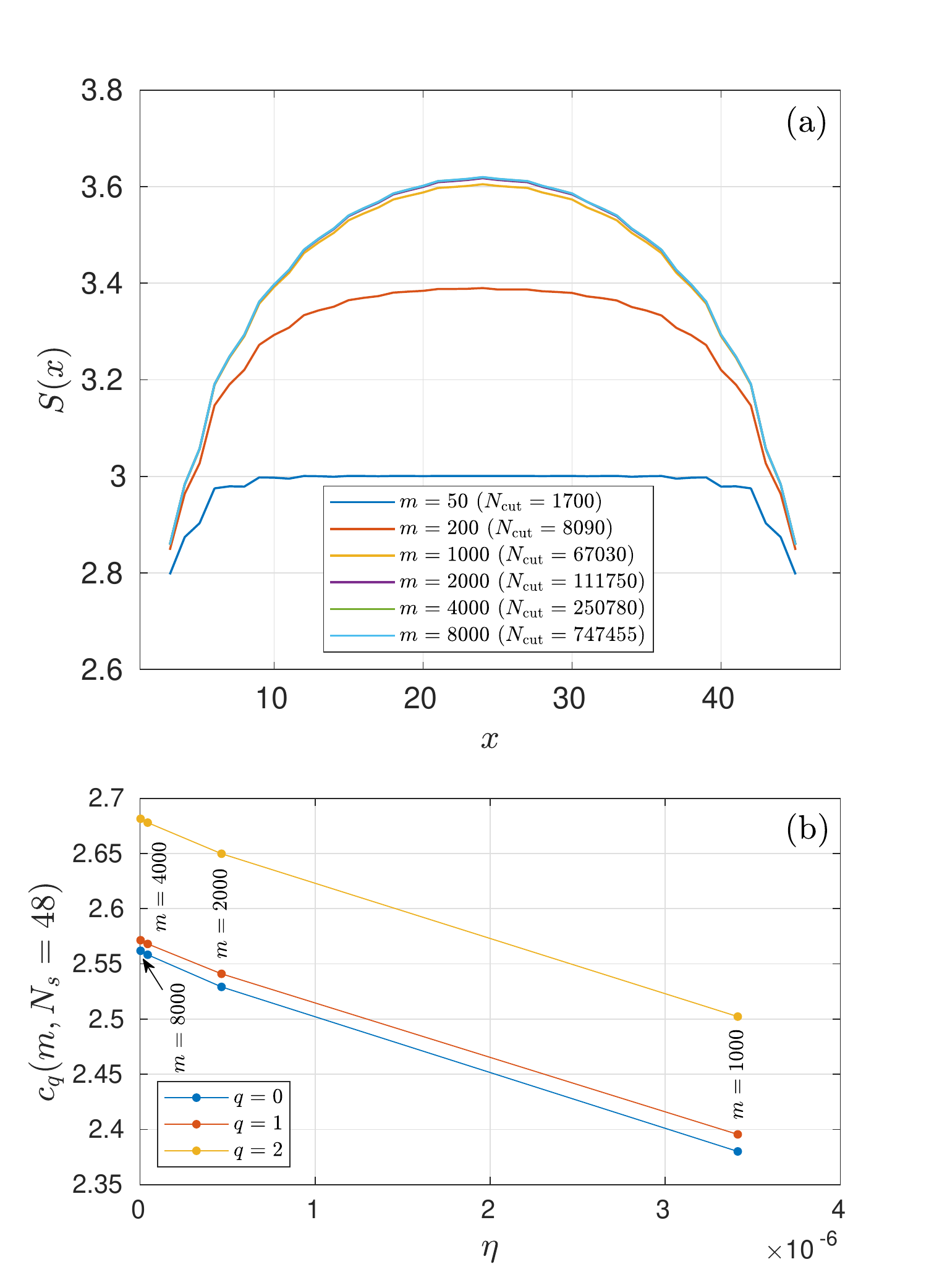}
\end{center}
\caption{(a) Entanglement entropy of a chain with $N_s=48$ sites. The value $N_{\rm cut}$ is the total number of states in the full Hilbert space that would be required to reach the same accuracy with a code which did not have the $\SU{N}$ symmetry. (b) Extrapolation of the central charge with respect to the discarded weight for $N_s = 48$ sites.}
\label{fig::supp:EE_and_extrapol_CC}
\end{figure}

The $3$-box symmetric $\SU{3}$ chain was claimed to be gapped based on the measurement of the entanglement entropy on a chain with $N_s=48$ sites with periodic boundary conditions, keeping $N_{\rm cut} = 1650$ states~\cite{rachel_spinon_2009}. Although we agree with the final conclusion (presence of a gap in the model), we show in Fig.~\ref{fig::supp:EE_and_extrapol_CC}(a) that it is not possible to reach such a conclusion on a short chain with an appropriate number of states kept.

Given the curvature of the entanglement entropy for small chains we fit the Calabrese-Cardy formula~\cite{calabrese_entanglement_2004},
\be
S(x) = \frac{c}{6} \ln\left( \frac{2 N_s}{\pi} \sin\left(\frac{\pi x}{N_s} \right) \right) + c_1
\ee
where $c$ is the central charge (for a critical model) and $c_1$ is a non-universal constant term, to our DMRG results as follows. For each value of the number of states kept $m$ we perform the fit on two points $\left\{ N_s - 3 - q, N_s-q\right\}$ separately for $q=0,1,2$ because of the oscillations appearing in the entanglement entropy along the chain. This defines the ``central charge'' $c_q(m,N_s)$ for finite $m$ and finite $N_s$. We then perform an extrapolation with respect to the discarded weight to obtain $c_q(m=\infty,N_s), \ q=0,1,2$ as shown in Fig.~\ref{fig::supp:EE_and_extrapol_CC}(b). The inset of Fig.~3 in the main text shows this quantity versus inverse chain length. For $N_s=48$ sites the extrapolated central charge lies between the ones of the $\SU{3}_1$ and $\SU{3}_2$ WZW conformal field theories. At $N_s=120$ sites the extrapolated central charge has become smaller than $c=2$, the smallest possible value for a critical model with $\SU{3}$ symmetry. Increasing further the system size the entanglement entropy ultimaltely becomes smaller than $c=1$ and seems to extrapolate to zero in the thermodynamic limit.

\section{Gap in the $\irrepconj$ sector}
\label{sec::330_heisenberg}

\begin{figure}
\begin{center}
\includegraphics[width=.5\textwidth]{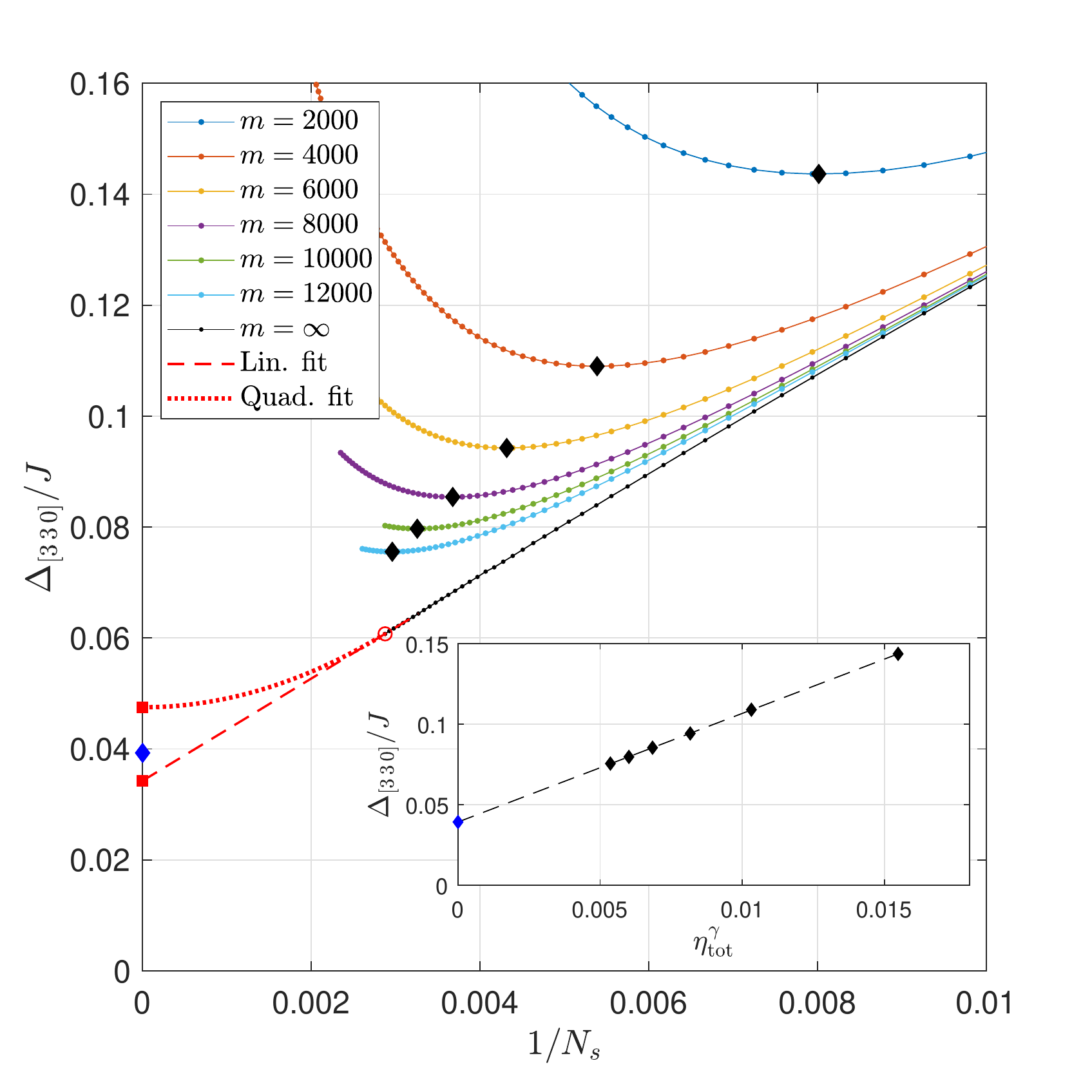}
\end{center}
\caption{Gap from the singlet sector to the $\irrepconj$ sector versus inverse chain length for different values of the number $m$ of states kept. The inset shows a power-law fit with exponent $\gamma = 0.46$ of the minimum of each finite-$m$ curve, denoted by a black diamond in the main plot, as a function of the total discarded weight $\eta_{\rm tot}$, which is dominated by the discarded weight in the $\irrepconj$ sector.}
\label{fig::330::gap_heisenberg}
\end{figure}

As mentioned in Sec.~\ref{sec::DMRG_alg} the DMRG simulations can also be performed in the conjugate sector $\irrepconj$. However the calculation of the reduced matrix elements of the interaction is significantly more demanding and we restrict ourselves to the first $M=102$ irreps. In Fig.~\ref{fig::330::gap_heisenberg} we show the analogue of Fig.~1 of the main text for the $\irrepconj$ irrep. Following the same procedure to extract the gap in the thermodynamic limit we obtain the approximate bounds $\Delta_{\irrepconj}/J \in \left[0.034, 0.048 \right]$ while the power-law fit with respect to the total discarded weight again provides a value in agreement with these bounds, $\Delta_{\irrepconj}/J \simeq 0.039$.

\section{Results at the AKLT point}
\label{sec::res_AKLT}

\begin{figure}
\begin{center}
\includegraphics[width=0.5\textwidth]{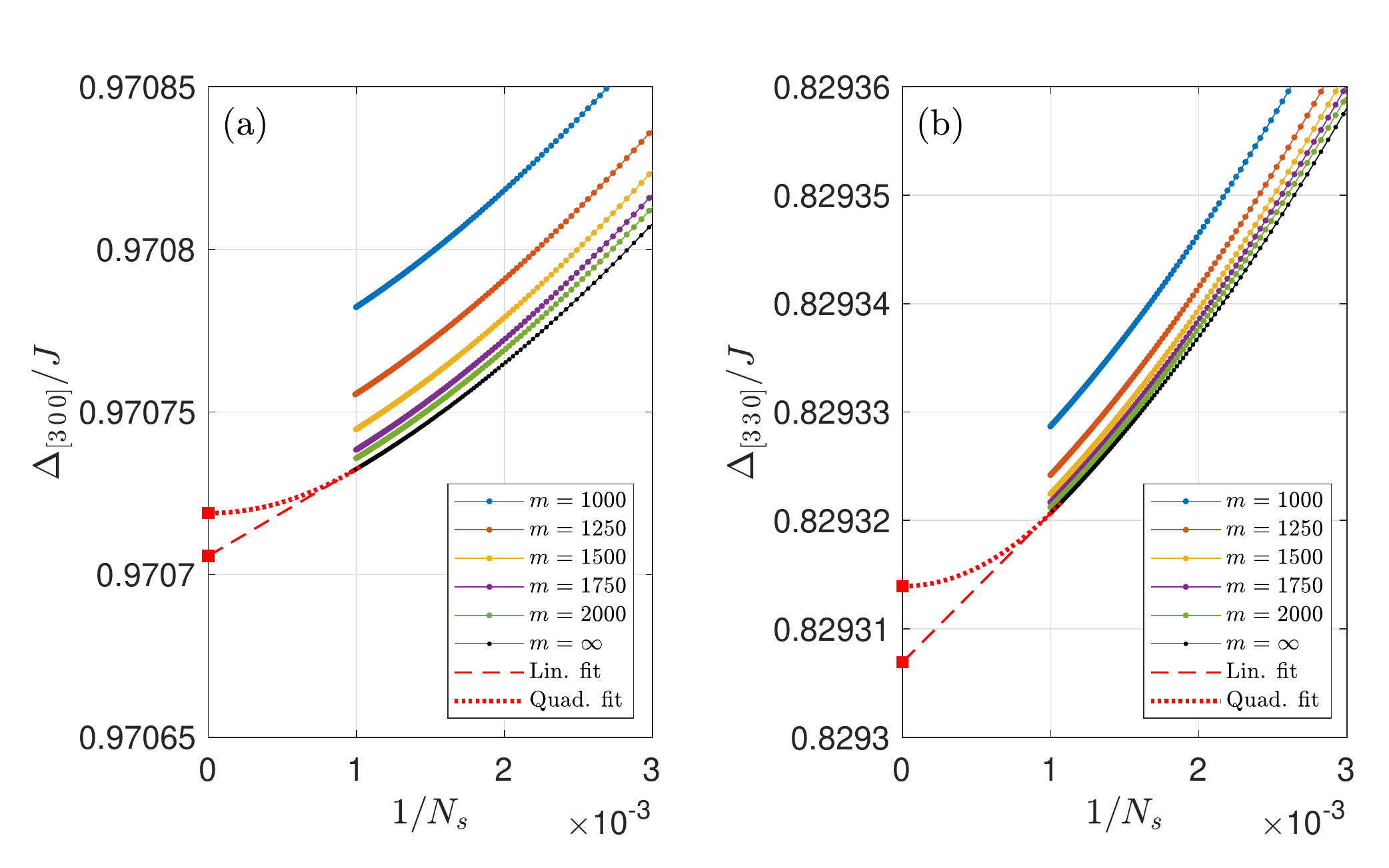}
\end{center}
\caption{Gaps at the AKLT point versus inverse chain length for different values of the number of states kept $m$. (a)~Singlet-$\irrepsymm$ gap. (b)~Singlet-$\irrepconj$ gap.}
\label{fig::SM::AKLT::gap300_and_gap330}
\end{figure}

In order to obtain Fig. 2 we have followed the same procedure as the one described in the main text for the Heisenberg model along the interpolation line between the AKLT Hamiltonian and the pure Heisenberg Hamiltonian. At the AKLT point the numerical calculation needs only to be performed in the symmetric sector $\irrepsymm$, not in the singlet sector. Figure~\ref{fig::SM::AKLT::gap300_and_gap330}(a) shows the gap from the singlet sector to the $\irrepsymm$ sector. As a consequence of the very short correlation length we obtain a very accurate value of the gap in the thermodynamic limit, $\Delta_{\irrepsymm}/J \in \left[ 0.970705, 0.970719\right]$, with a rather small number of states kept. To illustrate the picture of the AKLT state given in Ref.~\cite{gozel_novel_2019} we have also computed the entanglement entropy in the ground state of Hamiltonian~\eqref{equ::supp::aklt_H_final}. We obtain a completely flat entanglement entropy, with $S(x) = 1/\ln(8)$. This is precisely the hallmark of a valence bond solid state made of ``virtual'' $8$-dimensional adjoint representations.

In view of the monotonic decrease of the $\irrepsymm$ gap in Fig. 2, we claim that the AKLT and Heisenberg models lie in the same phase. We thus use the AKLT model, for which the correlation length is very short, to analyze the spectrum in different irreps. As explained in the main text, there are four Haldane gaps $\Delta_{\alpha}$ for the irreps $\alpha = \irrepadj, \, \irrepsymm, \, \irrepconj, \, [4\,2\,0]$, which is all the content appearing in the tensor product of two adjoint irreps (in addition to the singlet irrep):

\begin{center}
\begin{tikzpicture}
\draw (0,0) node[anchor=west]{$\shsc{2pt}{0.4}{\ydiagram{2,1}} \otimes \shsc{2pt}{0.4}{\ydiagram{2,1}} \ = \ \bullet \, \oplus 2 \shsc{2pt}{0.4}{\ydiagram{2,1}} \oplus \shsc{0pt}{0.4}{\ydiagram{3}} \oplus \shsc{2pt}{0.4}{\ydiagram{3,3}} \oplus \shsc{2pt}{0.4}{\ydiagram{4,2}}$};
\end{tikzpicture}
\end{center}

\begin{figure}
\begin{center}
\includegraphics[width=0.5\textwidth]{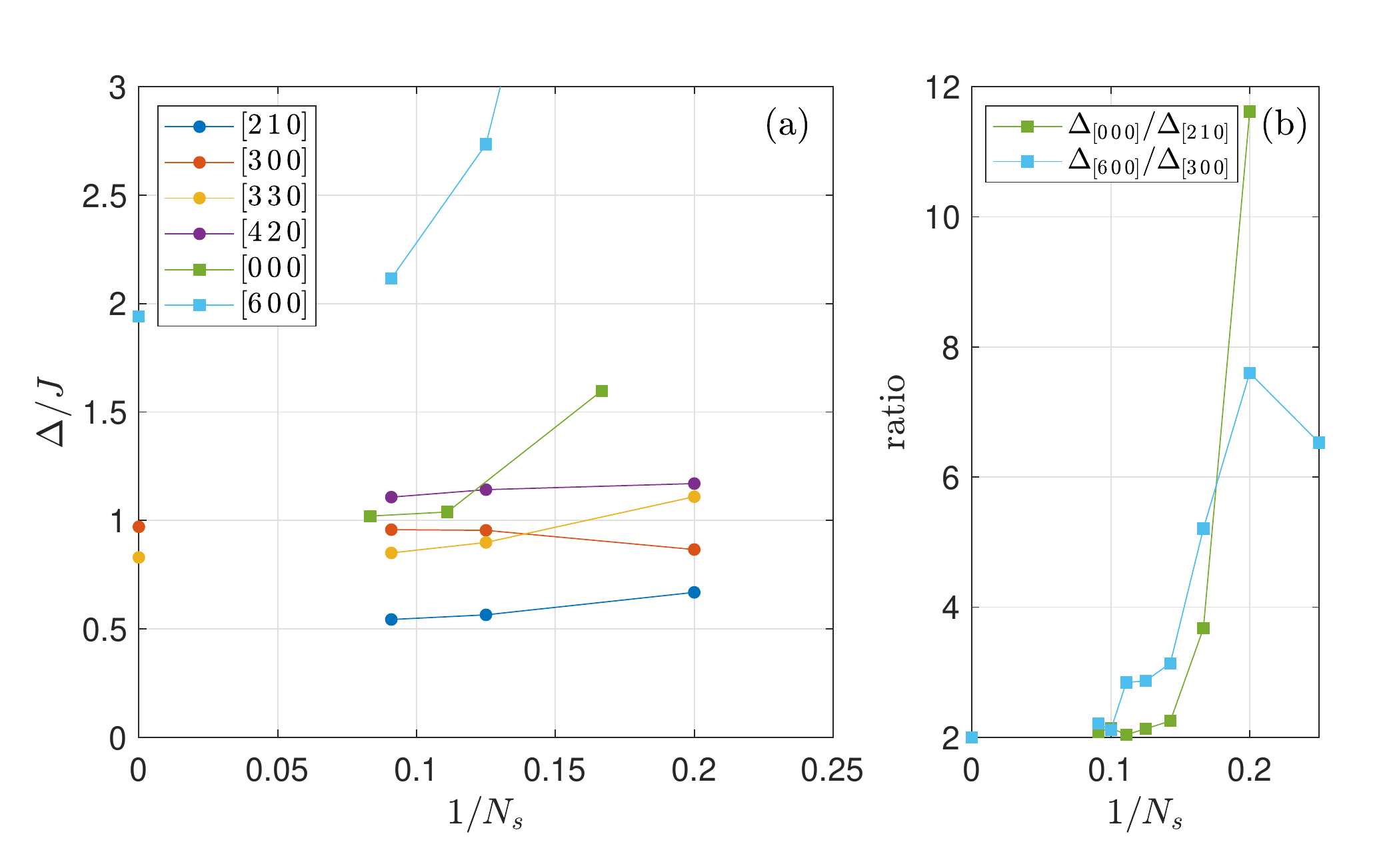}
\end{center}
\caption{(a) Finite-size gaps of the $\SU{3}$ AKLT Hamiltonian of Greiter and Rachel with PBC obtained by ED. The values at $1/N_s=0$ denote the thermodynamic gaps obtained by DMRG with OBC. Filled circles denote elementary excitations. Filled squares denote composite excitations. Lines are guides to the eye. (b) Ratio of the singlet gap versus the adjoint gap and the $\irrepsix$ gap versus the $\irrepsymm$ gap. Both ratios tend to $2$ in the thermodynamic limit, showing that the first singlet and $\irrepsix$ excitations are composite scattering excitations made of two adjoint, respectively $\irrepsymm$, elementary excitations.}
\label{fig::SM::gap_AKLT_GR_PBC}
\end{figure}

Figure~\ref{fig::SM::gap_AKLT_GR_PBC}(a) shows the gap to the first excited state in each of these sectors for the AKLT Hamiltonian with periodic boundary conditions for different chain lengths accessible with ED, as well as the singlet and $\irrepsix$ gaps. From these results we expect the adjoint gap $\Delta_{\irrepadj}$ to be the ${\it smallest}$ Haldane gap of the model, and to be approximately twice as small as $\Delta_{\irrepsymm}$. We further verify by DMRG at the AKLT point that the conjugate sector $\irrepconj$ is gapped. We obtain an extrapolated gap in the thermodynamic limit $\Delta_{\irrepconj}/J \in \left[ 0.829306, 0.829314 \right]$ (see Fig.~\ref{fig::SM::AKLT::gap300_and_gap330}(b)), smaller than the symmetric gap $\Delta_{\irrepsymm}$.

\begin{figure}
\begin{center}
\includegraphics[width=.5\textwidth]{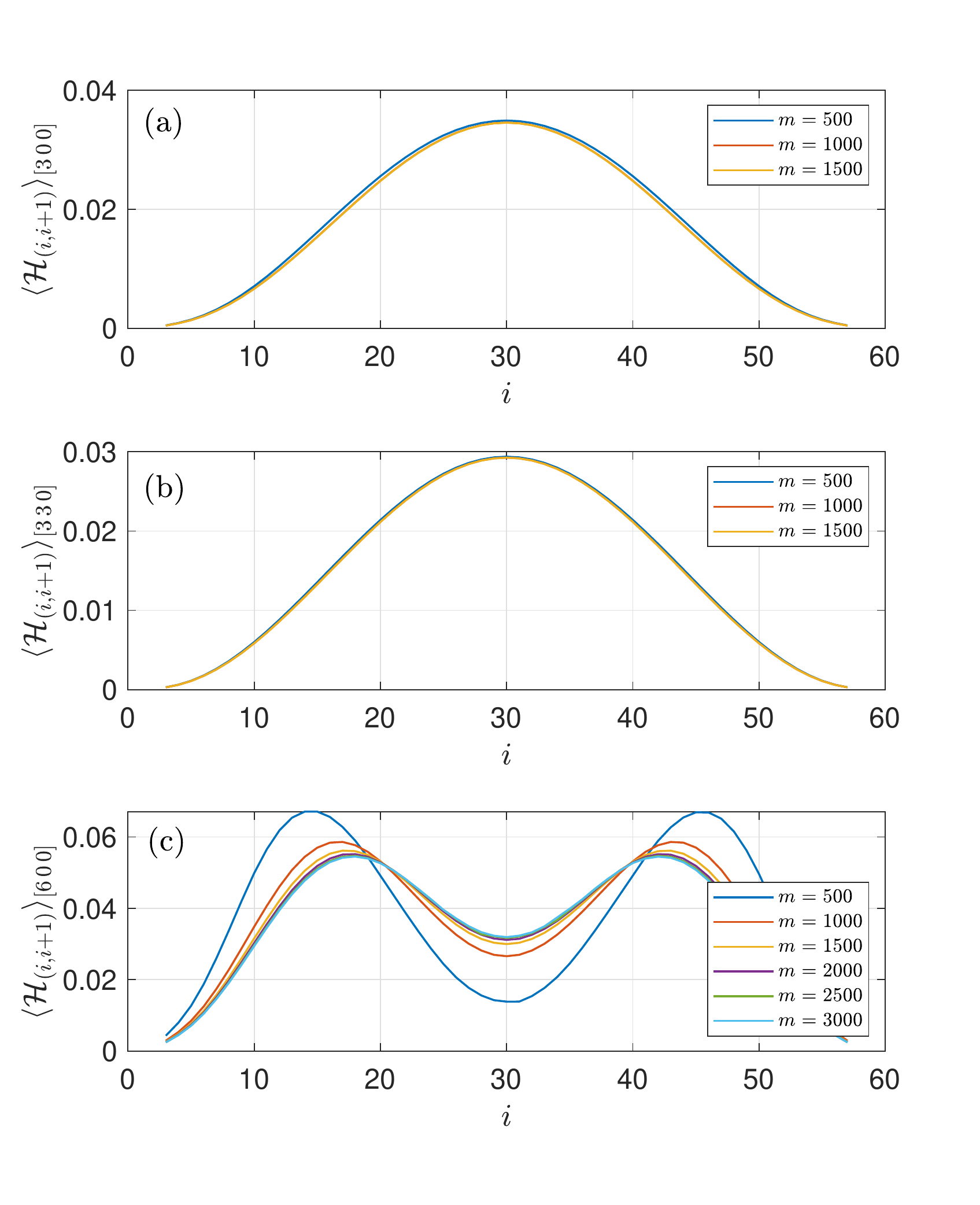}
\end{center}
\caption{Bond energy along the $\SU{3}$ chain with $N_s=60$ sites at the AKLT point in (a) the symmetric sector $\irrepsymm$, (b) the conjugate sector $\irrepconj$ and (c) the $6$-box symmetric sector $\irrepsix$. In all three plots we have subtracted the contribution of the singlet sector which is completely flat. We do not grant full convergence to the results in (c) because we are limited to the first $M=76$ irreps in this sector. The truncation error thus also depends on $M$.}
\label{fig::SM::AKLT::bond_energy}
\end{figure}

\begin{figure}
\begin{center}
\includegraphics[width=.5\textwidth]{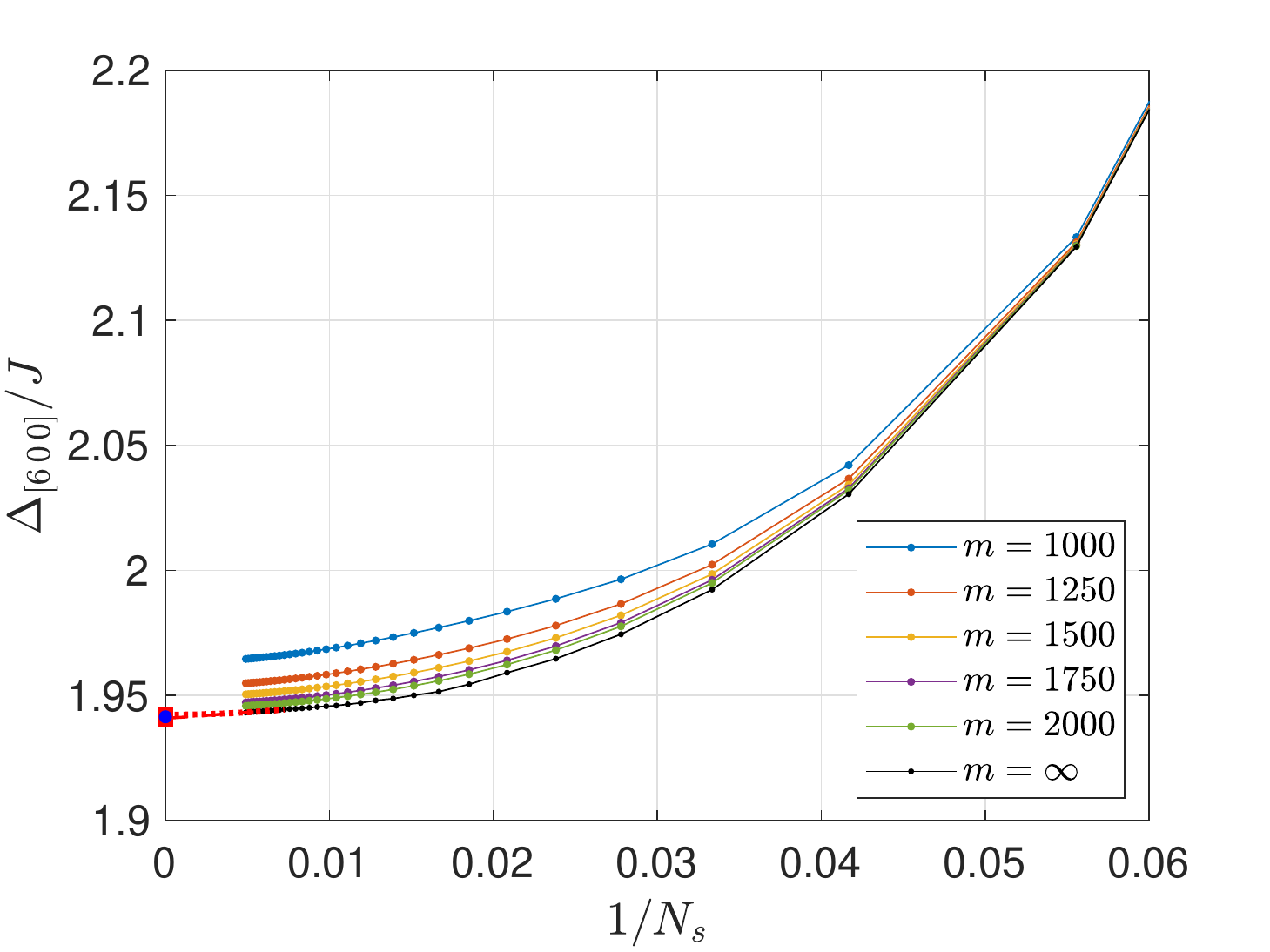}
\end{center}
\caption{Finite-size gap from the singlet ground state to the first $\irrepsix$ excited state at the AKLT point versus inverse chain length. A blue dot at $1/N_s=0$ is shown at energy $2\Delta_{\irrepsymm}$.}
\label{fig::SM::AKLT::gap_600_dmrg}
\end{figure}

In Fig.~\ref{fig::SM::AKLT::bond_energy}(a-b), following S\o{}rensen and Affleck in Ref.~\cite{sorensen_large_1993}, we show that the $\irrepsymm$ and $\irrepconj$ excitations are indeed elementary excitations, not composite excitations obtained by combining two or more other excitations, by computing the bond energy along the chain. The single peak structure is similar to the $1$-magnon state in a spin-$1$ chain (see Sec.~\ref{sec::su2} and Ref.~\cite{sorensen_large_1993}). Finally, we have investigated the gap and the bond energy along the chain in the $6$-box symmetric sector $\irrepsix$. This irrep can be accessed from the tensor product of two $\irrepsymm$ irreps:
\begin{center}
\begin{tikzpicture}
\draw (0,0) node[anchor=west]{$\shsc{0pt}{0.4}{\ydiagram{3}} \otimes \shsc{0pt}{0.4}{\ydiagram{3}} \ = \ \shsc{2pt}{0.4}{\ydiagram{3,3}} \oplus \shsc{2pt}{0.4}{\ydiagram{4,2}} \oplus \shsc{2pt}{0.4}{\ydiagram{5,1}} \oplus \shsc{0pt}{0.4}{\ydiagram{6}}$};
\end{tikzpicture}
\end{center}
or of two $[4\,2\,0]$ irreps:

\begin{center}
\begin{tikzpicture}
\draw (0,0) node[anchor=west]{$\shsc{2pt}{0.4}{\ydiagram{4,2}} \otimes \shsc{2pt}{0.4}{\ydiagram{4,2}} \ = \ \bullet \, \oplus 2 \shsc{2pt}{0.4}{\ydiagram{2,1}} \oplus \shsc{0pt}{0.4}{\ydiagram{3}} \oplus \shsc{2pt}{0.4}{\ydiagram{3,3}}$};
\draw (3,-1) node[anchor=west]{$\oplus \, 3 \, \shsc{2pt}{0.4}{\ydiagram{4,2}} \oplus \shsc{0pt}{0.4}{\ydiagram{6}} \oplus \, ...$};
\end{tikzpicture}
\end{center}
The bond energy, shown in Fig.~\ref{fig::SM::AKLT::bond_energy}(c) shows a double peak structure which is the signature of two elementary excitations which repel each other, exactly like a $2$-magnon state in the spin-$1$ chain (see Sec.~\ref{sec::su2} and Ref.~\cite{sorensen_large_1993}). The convergence of the results is not yet achieved because we are limited to the $M = 76$ first irreps of lowest quadratic Casimir. The calculation of the reduced matrix elements of the interaction for these shapes is very challenging as it took over $10^5$ CPU-hours. The singlet-$\irrepsix$ gap confirms the picture of two $\irrepsymm$ excitations which repel each other. We indeed obtain $\Delta_{\irrepsix}/J = 1.941(1)$, very accurately twice the singlet-$\irrepsymm$ gap $\Delta_{\irrepsymm}$, see Fig.~\ref{fig::SM::AKLT::gap_600_dmrg} and also Fig.~\ref{fig::SM::gap_AKLT_GR_PBC}(b).

Finally let us discuss shortly the adjoint and $[4\,2\,0]$ gaps, which we cannot compute with our DMRG algorithm. Thanks to the previous considerations however, one can deduce some inequalities, or bounds, that these gaps have to satisfy. In particular one has
\begin{equation}
\Delta_{\alpha} \geqslant \frac{1}{2} \max\left\{ \Delta_{\irrepsymm},\Delta_{\irrepconj} \right\}, \quad \alpha=\irrepadj, [4\,2\,0].
\end{equation}
This means that, at the Heisenberg point, the smallest bulk gap has to satisfy $\Delta_{\irrepadj}/J > 0.017$.

\section{Benchmark with the $\SU{2}$ spin-$1$ chain}
\label{sec::su2}

\begin{figure}[h]
\begin{center}
\includegraphics[width=0.5\textwidth]{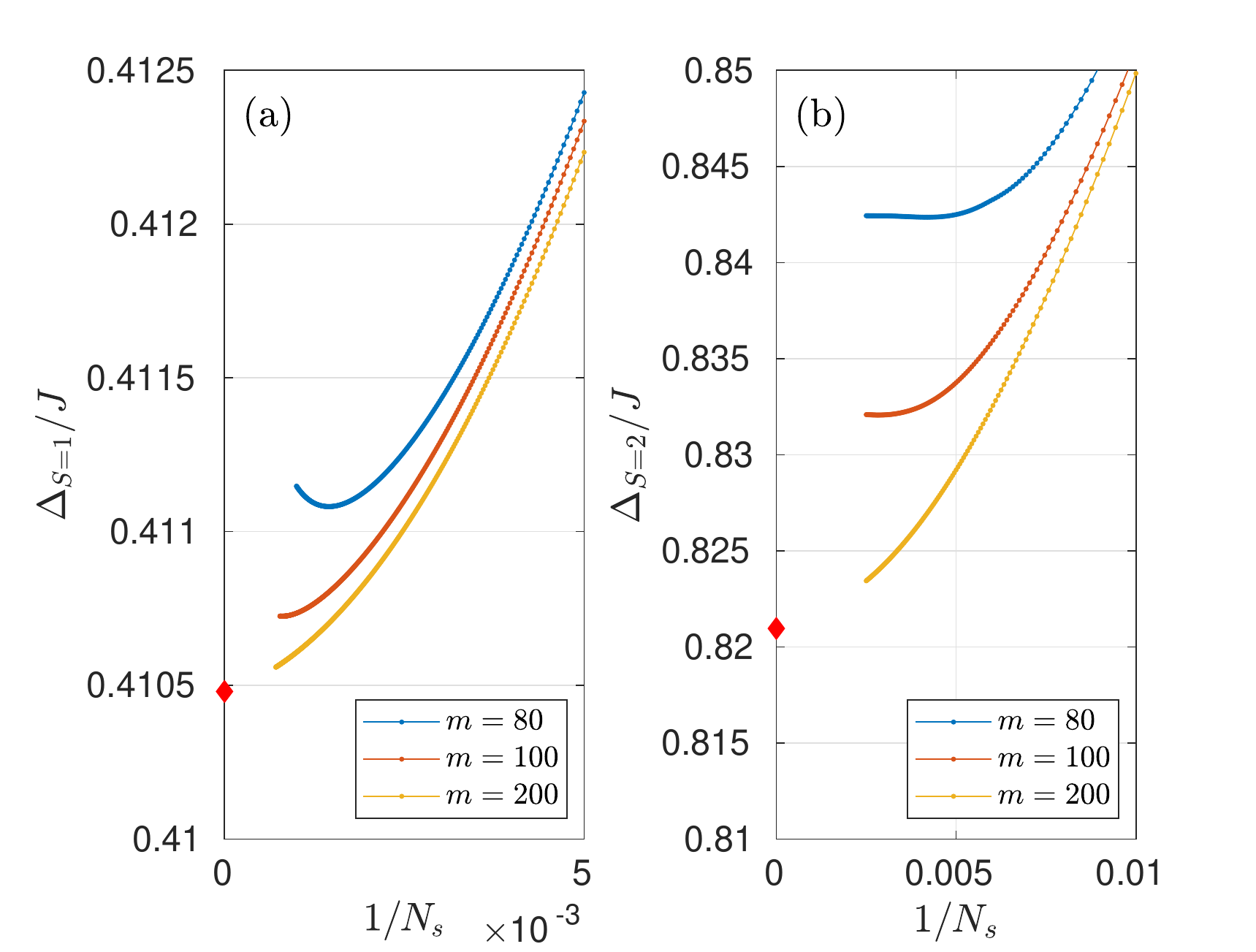}
\end{center}
\caption{Gap from the singlet ground state to (a) the first spin-$1$ excited state and (b) the first spin-$2$ excited state. The exact values $\Delta = 0.41048 J$ and $2\Delta$ are shown with a red diamond.}
\label{fig::sup::su2::gap}
\end{figure}

\begin{figure}[h]
\begin{center}
\includegraphics[width=0.5\textwidth]{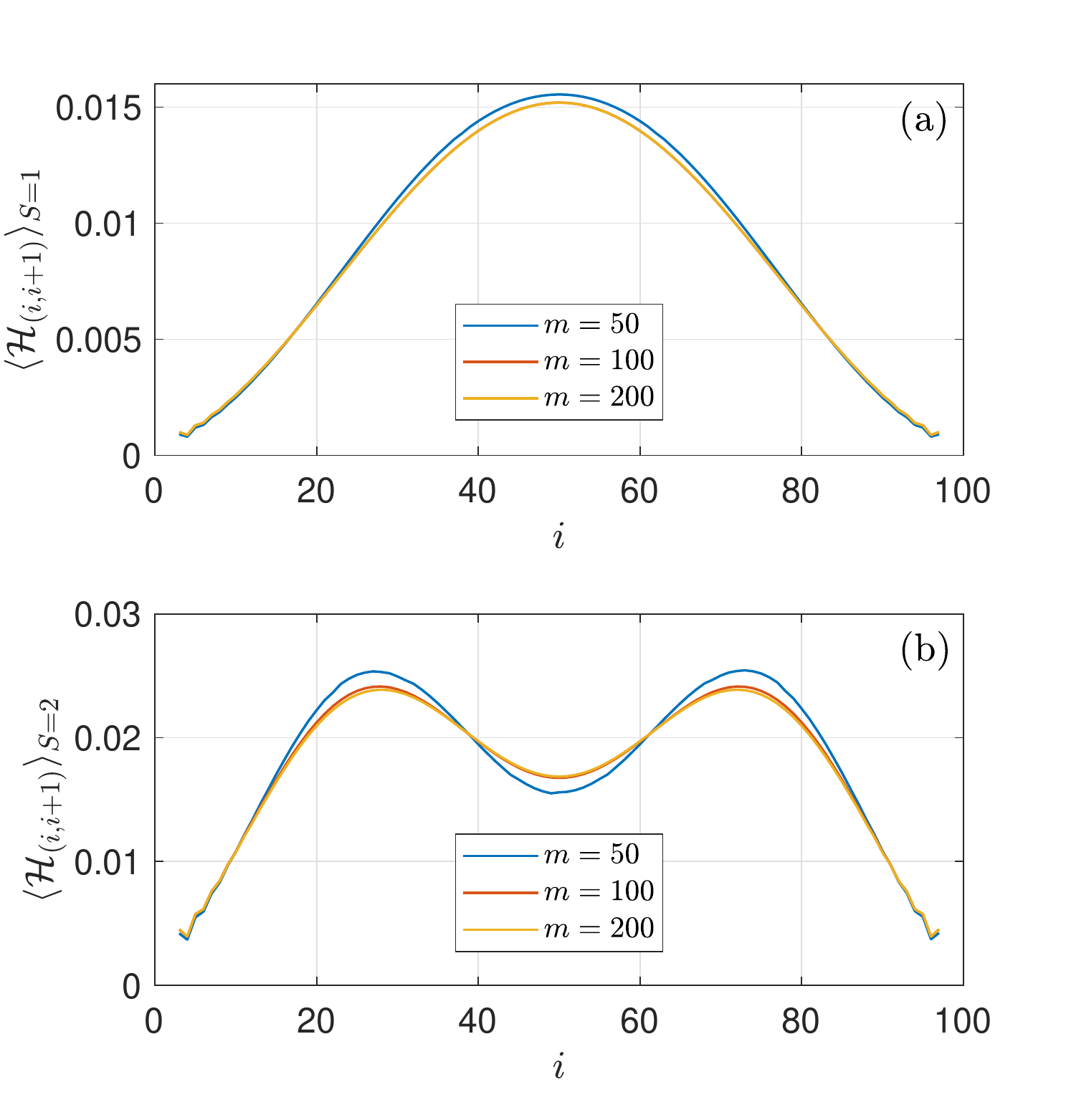}
\end{center}
\caption{Bond energy along the $\SU{2}$ spin-$1$ Heisenberg chain with the screening of the spin-$1/2$ edge states and with $N_s=100$ sites in (a) the spin-$1$ sector and (b) the spin-$2$ sector. In both figures we have subtracted the contribution of the ground state, which belongs to the singlet sector.}
\label{fig::sup::su2::correlation_S1_S2}
\end{figure}

The spectrum of the $\SU{2}$ spin-$1$ chain is well known: for an even numbered open chain where the edge states have not been screened, the ground state is a singlet and the first excited state is a low lying triplet which becomes degenerate with the singlet ground state exponentially fast as the system size increases. The Haldane gap $\Delta = 0.41048 J$ of the system is then found between the singlet ground state and the first spin-$2$ excited state. When the edge states are screened with spin-$1/2$ degrees of freedom the first spin-$1$ excited state is separated from the singlet ground state by the finite gap $\Delta$ and the first spin-$2$ excited state appears at energy $2\Delta$. It has been shown that this spin-$2$ excitation corresponds to two spin-$1$ excitations which repel each other~\cite{sorensen_large_1993}. In Fig.~\ref{fig::sup::su2::gap} we show the spin-$1$ and spin-$2$ gaps of the spin-$1$ chain with the screening of the spin-$1/2$ edge modes, obtained with our DMRG algorithm with full $\SU{2}$ symmetry. The gap from the singlet to the first spin-$2$ excitation indeed corresponds to twice the Haldane gap. In Fig.~\ref{fig::sup::su2::correlation_S1_S2} we present the bond energy along the chain in the spin-$1$ and spin-$2$ sectors. The double peak structure in Fig.~\ref{fig::sup::su2::correlation_S1_S2}(b) is the signature of the two spin-$1$ excitations forming a spin-$2$ state.

\clearpage

\nocite{*}
\bibliographystyle{apsrev4-1}
%